\documentclass[aps,prb,twocolumn,amsmath,amssymb,nofootinbib,eqsecnum,superscriptaddress,floatfix]{revtex4}

\usepackage{amsmath}
\usepackage{amssymb}
\usepackage[dvips]{color} 
\usepackage{graphicx}
\usepackage{dcolumn} 
\usepackage{bm} 
\usepackage[hypertex]{hyperref}
\usepackage{longtable}
\usepackage{ulem}

\newcommand{\Wick}[1]{\;\,\rule[10pt]{.5pt}{2pt}\!\hspace{1.2pt}\rule[12pt]{#1}{.5pt}
  \!\hspace{1.1pt}\rule[10pt]{.5pt}{2pt}\hspace{-#1}\hspace{-1.4mm}}

\newcommand{\Wickunder}[1]{\;\,\rule[-6pt]{.5pt}{2pt}\!\hspace{1.2pt}\rule[-6pt]{#1}{.5pt}
  \!\hspace{1.1pt}\rule[-6pt]{.5pt}{2pt}\hspace{-#1}\hspace{-1.4mm}}

\begin{document}


\title{ 
High-gradient 
operators in perturbed
Wess-Zumino-Witten
field theories in two dimensions
      }

\author{S.\ Ryu} 
\affiliation{
  Department of Physics, 
  University of California, Berkeley, CA 94720, USA
            } 
\author{C.\ Mudry} 
\affiliation{
  Condensed matter theory group, 
  Paul Scherrer Institute, CH-5232 Villigen PSI,
  Switzerland
            } 
\author{A.\ W.\ W.\ Ludwig} 
\affiliation{
  Department of Physics, 
  University of California, Santa Barbara, CA 93106, USA}
\author{A.\ Furusaki
            } 
\affiliation{
  Condensed Matter Theory Laboratory,
  RIKEN, Wako, Saitama 351-0198, Japan
            }

\date{\today}

\begin{abstract}
  Many classes of non-linear sigma models (NL$\sigma$Ms) 
  are known to contain composite operators with an arbitrary 
  number $2s$ of derivatives (``high-gradient operators'') 
  which appear to become strongly relevant within
  renormalization group (RG) calculations 
  at one (or fixed higher)
  loop order, when the number $2s$ of derivatives  becomes large.
  This occurs at many
  conventional fixed points of NL$\sigma$Ms which are perturbatively 
  accessible within the usual $\epsilon$-expansion in $d=2+\epsilon$ 
  dimensions. Since such operators are not
  prohibited from occurring in the action, they appear to threaten 
  the very existence of such fixed points.
  At the same time, for NL$\sigma$Ms describing  metal-insulator transitions
  of Anderson localization in electronic conductors, 
  the strong  RG-relevance of
  these operators has been previously related to statistical properties 
  of the conductance of samples of large finite size 
  (``conductance fluctuations'').
  In this paper, we analyze this question, 
  not for perturbative RG treatments of
  NL$\sigma$Ms, but for two-dimensional Wess-Zumino-Witten (WZW) models 
  at level $k$, 
  perturbatively in the current-current interaction of the Noether current 
  (``non-Abelian Thirring/Gross-Neveu models'').
  WZW models are special (``Principal Chiral'') 
  NL$\sigma$Ms on a Lie Group $G$ with  a WZW term at level $k$.  
  In these models the role of high-gradient operators is played by 
  homogeneous polynomials of order $2s$ in the Noether currents,
  whose scaling dimensions we analyze.
  For the Lie Supergroup $G=\mathrm{GL}(2N|2N)$ and $k=1$, this corresponds  
  to time-reversal invariant
  problems of Anderson localization  in the so-called chiral symmetry
  classes, and the strength of the
  current-current interaction, a measure of the strength of 
  disorder, is known to be completely marginal (for any $k$).
  We find that all high-gradient 
  (polynomial) operators are, to one loop order, 
  irrelevant or relevant depending on the 
  sign of that interaction.
\end{abstract}

\maketitle

\section{Introduction}
\label{sec: intro}

Fluctuations are known to play a key role in sufficiently
low-dimensional systems, whether classical or quantum, as they
can preempt spontaneous symmetry breaking.
When the symmetry is both global and continuous, 
the tool of choice to address the role
of fluctuations in low-dimensional systems is the non-linear sigma model
(NL$\sigma$M). However, the usefulness of NL$\sigma$Ms 
has come to transcend situations in which 
a pattern of symmetry breaking is immediately obvious.
For example, NL$\sigma$Ms
have been used with success in the context of Anderson localization
(see Ref.~\onlinecite{Efetov97} for a review)
to access the transition from a metallic to an insulating phase
induced by weak disorder or to compute probability distributions 
of spectral,\cite{Mirlin00} wavefunction,\cite{Foster09}
and transport characteristics in chaotic metallic grains and
disordered electronic systems.\cite{Efetov97}

Quite generally, the construction of a generic NL$\sigma$M on 
a connected Riemannian manifold $\mathfrak{M}$
of finite dimension $\mathfrak{n}$,
the ``target manifold'',
can proceed in the following way.%
\cite{Friedan85}
One assigns to any point from Euclidean space in $d$ dimensions,
specified by coordinates $x^{\mu}$ ($\mu=1,\cdots,d$),
a point in the manifold $\mathfrak{M}$ with the coordinates $\phi^{i}(x)$
($i=1,\cdots,\mathfrak{n}$).
The simplest action $S$, which is 
made of two derivatives of the coordinates $\phi^i$,
and is invariant under both the rotations of Euclidean space
and reparametrization of the target manifold, is
\begin{eqnarray}
\label{eq: construction NLSM on rieman manifold}
S=
\frac{1}{4\pi t}
\int \frac{d^{d}x}{\mathfrak{a}^{d-2}}
G^{\ }_{ij}\big[\phi(x)\big]
\partial^{\ }_{\mu}\phi^{i}(x)
\partial^{\ }_{\mu}\phi^{j}(x)
\end{eqnarray}
where $G^{ }_{ij}\left[\phi\right]$
is a component of the metric tensor on $\mathfrak{M}$,
$t$ is the coupling constant,
and $\mathfrak{a}$ is the short-distance cutoff.

The target manifold can be either 
compact or non-compact.
An example of a NL$\sigma$M on a compact target manifold 
is the O($N$)/O($N-1$) NL$\sigma$M 
with $2<N=3,4,5,\cdots$ when
the target manifold is the unit sphere $S^{N-1}$
in $N$-dimensional Euclidean space.
When $N=3$ it describes spontaneous symmetry breaking
in a classical ferromagnet.
Non-compact target manifolds 
are of relevance to the problem of Anderson localization 
in the  bosonic ``replica limit'' $N\to 0$ or
when the manifold is generalized to a supermanifold.%
\cite{Efetov97}
In Anderson localization the coupling constant $t$ has the meaning
of the inverse of the mean dimensionless conductance.%
\cite{Lee85}

The implicit assumption made in the construction%
~(\ref{eq: construction NLSM on rieman manifold})
is that all the invariant
scalars that contain $2s$ ($1<s=2,3,\cdots$) 
derivatives of the field can be ignored.
The standard justification for this assumption is that 
their ``engineering dimension'' $2s$ is
much larger than the spatial dimension $d=2+\epsilon$, 
i.e., they are irrelevant in the renormalization group (RG)
sense, and this is expected to remain so after renormalization in
$d=2+\epsilon$ dimensions 
for small $\epsilon$, and thus small $t$.

This assumption was called into question in Refs.%
~\onlinecite{Kravtsov88}--\onlinecite{Ryu07a},
for which the main results can be illustrated
most simply by the example
of the O($N$)/O($N-1$) NL$\sigma$M. We recall that the 
O($N$)/O($N-1$) NL$\sigma$M has
an 
infra-red 
unstable 
fixed point
located, to one loop order,
at $t^*=\epsilon/(N-2)$,
from which emerges a renormalization group (RG) flow
to strong and weak coupling.
In Ref.\ \onlinecite{Wegner90},
a family of perturbations of the 
O($N$)/O($N-1$) NL$\sigma$M action
(\ref{eq: construction NLSM on rieman manifold}),
which we shall call high-gradient operators, was considered.
A high-gradient operator of order $s$ is 
a homogeneous polynomial of order $2s$ in the derivatives of the fields
(all located at the same point) which is a scalar with respect
to both the symmetry group of the NL$\sigma$M [i.e., O($N$)]
and the rotation group of Euclidean space.
The minimum (i.e., dominant, or ``leading'') 
value of the one-loop scaling dimensions%
~\cite{footnote: our conventions for dimensions} 
of the high-gradient operators of order $s$
at the fixed point $t^{*}$ is found\cite{Wegner90} to be
\begin{eqnarray}
x^{(s)}&=&
2s 
- 
s(s-1)t^*
+
\mathcal{O}(\epsilon^2).
\label{eq: anomalous scaling dimensions for O(N) high-gradient operators}
\end{eqnarray}
Although strongly irrelevant by power counting 
(i.e., in the absence of fluctuation corrections, $t^* \to 0$),
high-gradient operators of
order $s$ thus acquire a one-loop scaling dimension smaller 
than two when the order $2s$ of derivatives
is large enough so that $s t^* \approx s \epsilon/(N-2) \sim 2$,
and thus would appear to become relevant,
based on the one-loop result.
In $d=2$ dimensions,
the lowest one-loop scaling dimension%
\cite{footnote: on anomalous dimensions}
for all high-gradient operators of order $s$ is
\begin{eqnarray}
x^{(s)}&=&
2s 
- 
s(s-1)t
+
\mathcal{O}(t^2)
\label{eq: anomalous dimensions for O(N) high-gradient operators}
\end{eqnarray}
along the trajectory to strong coupling 
away from the infra-red unstable fixed point $t=0$.
Two-loop counterparts to Eqs.%
~(\ref{eq: anomalous scaling dimensions for O(N) high-gradient operators})
and (\ref{eq: anomalous dimensions for O(N) high-gradient operators})
yield the same conclusion:~\cite{Castilla97}
high-gradient operators of sufficiently
high-order $s$ appear to be relevant for any given 
dimension $d=2+\epsilon$ at the non-trivial fixed point.

Similar results hold for the NL$\sigma$Ms defined on 
the compact target manifolds 
($M$ and $N$ are positive integers)
$\mathrm{Sp}(M+N)/\mathrm{Sp}(M)\times \mathrm{Sp}(N)$,%
~\cite{Kravtsov88,Kravtsov89}
$\mathrm{U}(M+N)/\mathrm{U}(M)\times \mathrm{U}(N)$,%
~\cite{Lerner90,Wegner91}
$\mathrm{O}(M+N)/\mathrm{O}(M)\times \mathrm{O}(N)$,%
~\cite{Mall93}
and on families of compact K\"ahler 
(and super) manifolds.%
~\cite{Ryu07a}
Generalizations to the non-compact target manifolds
$\mathrm{Sp}(M,N)/\mathrm{Sp}(M)\times\mathrm{Sp}(N)$,
$\mathrm{U}(M,N)/\mathrm{U}(M)\times\mathrm{U}(N)$,
and
$\mathrm{O}(M,N)/\mathrm{O}(M)\times\mathrm{O}(N)$
follow from the rule that the coupling $t$ of the 
compact NL$\sigma$M entering in one-loop anomalous dimensions
must be replaced by $-t$ in the corresponding non-compact
NL$\sigma$M. 
In Anderson localization, compact target manifolds arise
when using fermionic replicas for disorder averaging,
whereas non-compact target manifolds arise when using
the bosonic replicas for disorder averaging.
If one uses supersymmetric disorder averaging,
the resulting NL$\sigma$M has both compact 
and non-compact sectors.~\cite{Efetov97}
The high-gradient operators in the NL$\sigma$M defined on 
$\mathrm{AdS}_5\times S^5$
($\mathrm{AdS}_5$
is non-compact
whereas $S^5$
is compact)
have also been discussed 
in the context of 
the AdS/CFT correspondence.
(See, for example, Refs.%
~\onlinecite{Polyakov01}--\onlinecite{Polyakov05}.)

The substitution $t\to-t$ does not affect
the value of the minimal
(i.e., dominant, or ``leading'')
one-loop scaling dimensions, when the
spectrum of anomalous one-loop 
dimensions\cite{footnote: our conventions for dimensions} 
of all high gradient operators of order $s$ is
distributed symmetrically about zero.
This turns out to be the case 
whenever $m,n>1$ in the above examples.
On the other hand, there are some target manifolds, the simplest 
examples being 
$S^{N-1}=\mathrm{O}(N)/\mathrm{O}(N-1)$
and
$
\mathbb{C}P^{N-1}=
\mathrm{U}(N)/\mathrm{U}(N-1)\times \mathrm{U}(1)
$,
for which the full spectrum of one-loop anomalous dimensions
of order $s$ turns out to be not symmetric about zero,
in which case the substitution $t\to-t$ matters.
For example, high-gradient operators are made more irrelevant by one-loop
renormalization effects 
in the non-compact NL$\sigma$M on
$
\mathrm{U}(N-1,1)/\mathrm{U}(N-1)\times \mathrm{U}(1)
$.
(We refer the reader to 
Appendix~\ref{app: HGO on projective superspaces}
for a more detailed discussion of
``one-sided'' versus ``two-sided'' 
spectra of one-loop anomalous scaling dimensions 
for high-gradient operators in NL$\sigma$Ms.)

Of course, one can only conclude that high-gradient operators
become relevant for sufficiently large values of $s$,
if the strong relevance seen in the one-loop expressions for their
scaling dimensions persists when all higher loop contributions
(not computed here or in other works on this subject)
have been taking into account.
For example, the one-loop expressions may not be
characteristic in the large-$s$ limit, if the actual
expansion parameter is not $\epsilon$ but $s\epsilon$.%
\cite{%
Ludwig90,%
Brezin97%
     }
As any insight for resolving the nature of the $\epsilon$
expansion for high-gradient operators in NL$\sigma$Ms
must come from outside the $\epsilon$ expansion itself, 
progress has stalled since the early 1990's.

The aim of this paper is to study
the operators that play the role of the high-gradient operators
in field theories which are two-dimensional Wess-Zumino-Witten 
(WZW) theories~\cite{Wess71}$^{-}$\cite{Bocquet00}
on a Lie group $G$,
perturbed by 
an interaction quadratic in the Noether currents 
(``current-current interaction'').
Such theories are often also referred to as
``two-dimensional non-Abelian Thirring
(or Gross-Neveu) models''.
Any WZW theory, 
which is a Principal-Chiral-Non-Linear-sigma model
supplemented by a WZW term at level $k$,
gives a prescription to construct high-gradient operators
in terms of powers of Noether currents. 
Because it is possible to represent the Noether currents in WZW theories
in terms of free fermions,\cite{Witten84} 
one might be inclined to think
that such operators are perhaps not capable of 
displaying a ``pathological''
spectrum of scaling dimension
as in Eq.%
~(\ref{eq: anomalous scaling dimensions for O(N) high-gradient operators}).
However, as we demonstrate in this paper,
the situation is more interesting. Indeed, we will see
that under conditions specified below,
the one-loop spectra of the form
(\ref{eq: anomalous dimensions for O(N) high-gradient operators})
and
(\ref{eq: anomalous scaling dimensions for O(N) high-gradient operators})
can be realized by perturbing a WZW critical point by 
a current-current perturbation.

We also want to investigate if there is a difference 
between the properties of high-gradient operators
in unitary and non-unitary non-Abelian Thirring models.
This is important because
NL$\sigma$Ms describing the physics of
Anderson localization are non-unitary field theories.
Moreover,  high-gradient operators  in these theories
have been previously related to the statistical fluctuations
of the conductance of a disordered metal.%
~\cite{Altshuler86a}$^{-}$\cite{Altshuler91}
In this context, an appealing physical interpretation of the spectra
(\ref{eq: anomalous dimensions for O(N) high-gradient operators})
and
(\ref{eq: anomalous scaling dimensions for O(N) high-gradient operators})
has been proposed, attributing them to a broad tail
in the probability distribution of the conductance.%
~\cite{Altshuler86a}$^{-}$\cite{Altshuler91}
However, given that this interpretation 
depends crucially on the ability to invert
the $s\to\infty$ and $\epsilon\to0$
limits in spectra which are analogous to those in Eqs.%
~(\ref{eq: anomalous dimensions for O(N) high-gradient operators})
and
~(\ref{eq: anomalous scaling dimensions for O(N) high-gradient
operators}),
it would be useful to have
an example of a critical field theory
describing a problem of Anderson localization
for which one can study the RG-relevance of high-gradient operators 
without resorting to the $\epsilon$-expansion,
and for which one can reasonably expect a broad distribution 
of the conductance.

We now provide an outline of the article and a summary of our results.

It is shown in Sec.\ \ref{sec: su2} that high-gradient operators
in the (unitary) $\widehat{\mathrm{su}}(2)^{\ }_{k}$ Thirring model 
with strength $g$ of the ``current-current interaction'',
are made more irrelevant by the presence of these interactions
when the latter are (marginally) irrelevant
($g<0$, in our conventions).
On the other hand,
along the renormalization group (RG) flow driven
by a (marginally) relevant current-current interaction 
($g>0$, in our conventions),
a one-loop spectrum of the form%
~(\ref{eq: anomalous dimensions for O(N) high-gradient operators})
is recovered
in the ``classical'' limit $1/k\to0$.
The inverse level $1/k$ plays here
the role of a ``quantum'' parameter.
Indeed, for any finite $k$, we find that
the quadratic growth in $s$ in the 
unbounded one-loop spectrum 
~(\ref{eq: anomalous dimensions for O(N) high-gradient operators})
does not persist for values of $s$ larger than $k$.
In effect, $1/k$ determines the efficiency in ``taming'' the
strong RG-relevance of high-gradient operators
seen at one-loop order,
which is related to the fact that there exists a representation
of the current algebra of the level-$k$ 
WZW theory in terms of free fermions.

Section~\ref{sec: gl(M|N) level k=1}
is devoted to high-gradient operators in
what we will call the $\widehat{\mathrm{gl}}(M|M)^{\ }_{k}$ Thirring 
(or Gross-Neveu) model which was discussed in Ref.~\onlinecite{Guruswamy00}.
This is the $\widehat{\mathrm{gl}}(M|M)^{\ }_{k}$
WZW theory
on the Lie Supergroup $\mathrm{GL}(M|M)$,
perturbed by 
\textit{two}  current-current perturbations,
one which we call $g^{\ }_{\mathrm{M}}$, which
 is exactly marginal, and 
another
which we call $g^{\ }_{\mathrm{A}}$,
which flows logarithmically under the RG at a rate dependent on 
$g^{\ }_{\mathrm{M}}$.
In spite of the presence of an RG
flow of the coupling $g^{\ }_{\mathrm{A}}$
there exists a sector of the theory, the
so-called $\mathrm{PSL}(M|M)$ sector,
which is scale (conformally) invariant throughout.%
~\cite{Guruswamy00}
The high-gradient operators turn out to reside
in this conformally invariant sector,
and are unaffected by  the presence of the coupling $g^{\ }_{\mathrm{A}}$.
We will show that, for $k=1$,  the spectrum of one-loop anomalous
dimensions of high-gradient operators is fundamentally different
for positive and negative
values of the coupling constant
$g^{\ }_{\mathrm{M}}$.
In particular, when $g^{\ }_{\mathrm{M}}>0$ 
all high-gradient operators are made more irrelevant by the
current-current perturbations,
whereas  they are made more relevant when
$g^{\ }_{\mathrm{M}}<0$.
We close Sec.~\ref{sec: gl(M|N) level k=1}
by comparing the anomalous scaling dimensions
of high-gradient operators in
the $\widehat{\mathrm{gl}}(M|M)^{\ }_{k}$ Thirring 
(or Gross-Neveu) models and those in the 
$\mathrm{GL}(2N|2N)/\mathrm{OSp}(2N|2N)$
NL$\sigma$Ms, observing that they behave in the same way.

The result that the spectrum of one-loop anomalous dimensions of
high-gradient operators is strongly dependent on
the sign of $g^{\ }_{\mathrm{M}}$
has important implications  
in the context of Anderson localization because, as discussed in Ref.%
~\onlinecite{Guruswamy00},
the $\widehat{\mathrm{gl}}(M|M)^{\ }_{k}$ 
Thirring (or Gross-Neveu) model at $k=1$
describes a disordered electronic system, where
$g^{\ }_{\mathrm{A}} >0$ and $g^{\ }_{\mathrm{M}}>0$
correspond to the strengths of disorder potentials. 
The theory with $g^{\ }_{\mathrm{M}}>0$ thus offers
an example of a critical theory for Anderson localization with no
relevant high-gradient operator.
--  For example, this field theory describes~\cite{Guruswamy00}  
a tight-binding model of electrons on the honeycomb lattice 
with (real-valued) random hopping matrix elements
which are non-vanishing only between the two sublattices of the  
bipartite honeycomb lattice (see also Ref.~\onlinecite{Foster06}).
Versions of the honeycomb tight-binding model provide the basic electronic
structure of graphene.
In the classification scheme of Zirnbauer, and Altland and Zirnbauer,%
~\cite{Verbaarschot94}$^{-}$\cite{Heinzner05}
this model belongs to 
the ``chiral-orthogonal'' symmetry class (class BDI).
(Another example of a problem of Anderson localization 
in the same symmetry class 
is provided by a random tight-binding model on a square lattice
with $\pi$-flux through every plaquette.\cite{Hatsugai97})

By contrast, when  $g^{\ }_{\mathrm{M}}<0$,
the spectrum of one-loop scaling dimensions is unbounded
from below for any $k$ as is the case in Eq.%
~(\ref{eq: anomalous scaling dimensions for O(N) high-gradient operators}).
The full spectrum of one-loop anomalous scaling dimensions of
high gradient operators as it appears, e.g.,  
in the Grassmanian NL$\sigma$Ms with target manifolds
$\mathrm{Sp}(M+N)/\mathrm{Sp}(M)\times \mathrm{Sp}(N)$,
$\mathrm{U}(M+N)/\mathrm{U}(M)\times \mathrm{U}(N)$,
$\mathrm{O}(M+N)/\mathrm{O}(M)\times \mathrm{O}(N)$,
which is symmetric about zero,
is only recovered in the extreme ``classical'' limit
$M,k\to\infty$. In the context of Anderson localization,
the case with $g^{\ }_{\mathrm{M}}<0$
describes the surface state of a
three-dimensional topological insulator
in the chiral-symplectic class (symmetry class CII)
of Anderson localization.%
~\cite{Schnyder08,Ryu09,Hosur09}

After concluding in Sec.~\ref{sec: conclusion},
we review in Appendix~\ref{sec: HWK model}
the realization of the
$\widehat{\mathrm{gl}}(2N|2N)^{\ }_{k=1}$ 
Thirring (or Gross-Neveu) model as a problem
of Anderson localization in two dimensions
in symmetry class BDI, 
which was established in
Ref.~\onlinecite{Guruswamy00}.

\section{ 
High-gradient operators and
$\widehat{\mathrm{su}}(2)^{\ }_{k}$ WZW theories 
        }
\label{sec: su2}

The O(3)/O(2) NL$\sigma$M with coupling constant $t$
is the simplest example of a NL$\sigma$M 
containing \textit{infinitely many} high-gradient operators
all of which would appear to become 
relevant based on one-loop results.
This happens at the infra-red unstable
fixed point $t^{*}=\epsilon$ 
in $d=2+\epsilon >2 $ dimensions within the one-loop approximation
as long as the order $s$ of these high-gradient operators is
large enough. A precursor to this perturbative property
also occurs in $d=2$ dimensions close to the infra-red 
unstable fixed point $t=0$ 
as the NL$\sigma$M flows to strong coupling. Along this flow,
the spectrum of one-loop dimensions%
~\cite{footnote: on anomalous dimensions}
for the high-gradient operators 
is unbounded from below.

In two dimensions, 
the O(3)/O(2) NL$\sigma$M, supplemented by a topological
theta-term at $\theta=\pi$,
flows to a critical field theory, 
the SU(2) WZW theory with $\widehat{\mathrm{su}}(2)^{\ }_{k=1}$
current algebra,
$\widehat{\mathrm{su}}(2)^{\ }_{k=1}$
WZW theory.%
~\cite{Affleck87}$^{-}$\cite{Zamolodchikov92}
The strongly relevant high-gradient operators 
near the infra-red unstable fixed point $t=0$ 
must become
irrelevant at the WZW critical point, 
because the full operator content of the 
$\widehat{\mathrm{su}}(2)^{\ }_{k=1}$ WZW theory
is known to contain only a finite number 
of relevant 
fields (with scaling dimensions bounded from below
and above by zero and two, respectively).
The purpose of this section is to perturb the
$\widehat{\mathrm{su}}(2)^{\ }_{k}$ WZW theory
with a current-current perturbation
and to examine the fate of those operators in the
$\widehat{\mathrm{su}}(2)^{\ }_{k}$ WZW theory
which correspond to the
high-gradient operators in the O(3)/O(2) NL$\sigma$M.
We will refer to these operators still as ``high-gradient operators''.
We are going to argue that the spectrum of
one-loop scaling dimensions 
associated with
all high-gradient operators is bounded from below by the lowest
one-loop scaling dimension corresponding to
high-gradient operators of order $k$.
This result is very different from the unbounded
spectrum~(\ref{eq: anomalous dimensions for O(N) high-gradient operators})
of one-loop scaling dimensions associated with
high-gradient operators in the two-dimensional
O(3)/O(2) NL$\sigma$M.

In the following sections, we first review the
$\widehat{\mathrm{su}}(2)^{\ }_{k}$ WZW theory
perturbed by a current-current interaction.
Second, we identify high-gradient operators of order $s$.
Finally, we compute the leading one-loop dimensions of
high-gradient operators of order $s$ up to one loop.

\subsection{
Definitions
           }

The most fundamental property 
of the $\widehat{\mathrm{su}}(2)^{\ }_{k}$ WZW theory
is the existence of
a pair of holomorphic and antiholomorphic SU(2) Noether
currents, 
$J ^{\ }_1$,    
$J^{\ }_2$,     
$J^{\ }_3$,
and 
$\bar J^{\ }_1$,
$\bar J^{\ }_2$,
$\bar J^{\ }_3$,
respectively,
which satisfy the affine (Kac-Moody) current algebra
\begin{subequations}
\label{eq: su(2) level k current algebra}
\begin{equation}
\begin{split}
&
J^{\ }_{\alpha}(z) J^{\ }_{\beta}(0)=
\frac{k C^{\ }_{\alpha\beta} }{z^{2}}
+
\frac{{i}}{z}
 f^{\ }_{\alpha\beta}{}^{\gamma}
J^{\ }_{\gamma}(0)
+\cdots,
\\
&
\bar J^{\ }_{\alpha}(\bar z) \bar J^{\ }_{\beta}(0)=
\frac{kC^{\ }_{\alpha\beta}}{\bar z^{2}}
+
\frac{{i}}{\bar z}
 f^{\ }_{\alpha\beta}{}^{\gamma}
\bar J^{\ }_{\gamma}(0)
+\cdots,
\\
&
J^{\ }_{\alpha}(z)\bar J^{\ }_{\beta}(0)=
0,
\end{split}
\label{eq: su(2) level k current algebra a}
\end{equation}
at level $k=1,2,3,\cdots,$
where the invariant (Casimir) tensor 
of rank 2 in $\mathrm{su}(2)$
has the contravariant and covariant representations
(in our conventions)
\begin{equation}
C^{\ }_{\alpha\beta}=
\frac{1}{2}\delta^{\ }_{\alpha\beta},
\qquad
C^{\alpha\beta}=
2\delta^{\ }_{\alpha\beta},
\label{eq: invariant Casimir}
\end{equation}
respectively, while the structure constant of su(2) 
is 
the fully antisymmetric Levi-Civita tensor of rank 3,
\begin{eqnarray}
f^{\ }_{\alpha\beta}{}^{\gamma}=
\epsilon^{\ }_{\alpha\beta\gamma},
\qquad
\alpha,\beta,\gamma=1,2,3.
\label{eq: structure constants su(2)}
\end{eqnarray}
\end{subequations}
The dots in Eq.~(\ref{eq: su(2) level k current algebra a})
are terms of order zero and higher in powers of $z$ ($\bar z$) 
with $z=x+{i}y$ 
($\bar z=x-{i}y$)  
the holomorphic (antiholomorphic) coordinates
of the Euclidean plane. We shall also refer to the (anti) holomorphic sector 
of the theory as the (right-) left-moving sector.

The $\widehat{\mathrm{su}}(2)^{\ }_{k}$ current algebra%
~(\ref{eq: su(2) level k current algebra})
has a representation 
in terms of free-fermions. 
More precisely,
it is obtained from the action%
~\cite{footnote: Einstein convention}
\begin{subequations}
\label{eq: def free fermion rep of su(2) level k}
\begin{equation}
S^{\ }_{*}:=
\sum_{\iota=1}^{k}
\int \frac{d\bar zd z}{2\pi{i}}
\left(
    \psi^{c\dag}_{\ \iota}
\bar \partial\,
    \psi^{\ }_{c\iota}
+
\bar \psi^{c\dag}_{\ \iota}
     \partial\,
\bar \psi^{\ }_{c\iota}
\right) 
\label{eq: def S*}
\end{equation}
constructed from $k$-independent flavors of 
left ($\psi$) and right ($\bar\psi$) moving Dirac fermions,
whereby each one transforms 
in the fundamental representation of SU(2)$\times$SU($k$), 
with the partition function
\begin{equation}
Z^{\ }_{*}:=
\int\mathcal{D}[\psi^{\dag},\psi,\bar\psi^{\dag},\bar\psi]\,
\exp\left(-S^{\ }_{*}\right).
\label{eq: def Z*}
\end{equation}
\end{subequations}
One has the operator product expansions (OPE)
\begin{equation}
\begin{split}
&
    \psi^{\ }_{c\iota}(z)
    \psi^{d\dag}_{\ \iota'}(0)=
    \psi^{d\dag}_{\ \iota'}(z)
    \psi^{\ }_{c\iota}(0)
\sim
\frac{\delta^{\ }_{\iota\iota'}\delta^{\ }_{cd}}{     z},
\\
&
\bar\psi^{\ }_{c\iota}(z)
\bar\psi^{d\dag}_{\ \iota'}(0)=
\bar\psi^{d\dag}_{\ \iota'}(z)
\bar\psi^{\ }_{c\iota}(0)
\sim
\frac{\delta^{\ }_{\iota\iota'}\delta^{\ }_{cd}}{\bar z},
\\
&
\psi^{\ }_{c\iota}(z)
\bar \psi^{d\dag}_{\,\iota'}(0)
\sim
0,
\\
\end{split}
\label{eq: OPE free left and right movers}
\end{equation}
for $\iota,\iota'=1,\cdots,k$ and $c,d=1,2$.
In turn, the OPE~(\ref{eq: OPE free left and right movers})
imply that the left and right Noether currents
\begin{equation}
\begin{split}
     J^{\ }_{\alpha}:=
\sum_{\iota=1}^{k}
    \psi^{c\dag}_{\ \iota}
\frac{(\sigma^{\ }_{\alpha})^{\ }_{c}{}^{d}}{2}
    \psi^{\ }_{d\iota},
\quad
\bar J^{\ }_{\alpha}:=
\sum_{\iota=1}^{k}
\bar \psi^{c\dag}_{\ \iota}
\frac{(\sigma^{\ }_{\alpha})^{\ }_{c}{}^{d}}{2}
\bar\psi^{\ }_{d\iota},
\end{split}
\label{eq: fermionic oscillator rep of su(2)}
\end{equation}
with $\alpha=1,2,3$
obey the  $\mathrm{SU}(2)^{\ }_{k}$ 
current algebra~(\ref{eq: su(2) level k current algebra}).

The field theory defined by
Eq.~(\ref{eq: def free fermion rep of su(2) level k})
is a free-fermion field theory. 
The content of  local operators is thus known.
It contains a finite number of fields
whose scaling dimensions are bounded between 0 and 2
and are thus relevant, 
as it should be for a field theory defined on a Hilbert space with 
a positive definite inner product 
and with a spectrum bounded from below which is built on 
the Dirac-Fermi sea, in short a unitary field theory. 
Clearly, within the set of powers of the Noether currents
(\ref{eq: fermionic oscillator rep of su(2)}) there is thus
no room for an infinite family of relevant operators.

We perturb the free-fermion field theory
by a current-current interaction 
$\mathcal{O}^{\ }_I$
of the $\mathrm{SU}(2)^{\ }_{k}$ currents 
(\ref{eq: fermionic oscillator rep of su(2)}).

\begin{equation}
\begin{split}
&
Z:=
\int \mathcal{D}[\psi^{\dag},\psi,\bar\psi^{\dag},\bar\psi]\,
\exp\left(-S\right),
\\
&
S:=
S^{\ }_{*}
+
g 
\int\frac{d\bar zdz}{2\pi{i}} 
\mathcal{O}^{\ }_I(\bar z,z),
\\
&
\mathcal{O}^{\ }_I(\bar z,z):=
C^{\alpha\beta}J^{\ }_{\alpha}(z)\bar J^{\ }_{\beta}(\bar z)\equiv
2 J^{\ }_{\alpha}(z)\bar J^{\ }_{\alpha}(\bar z).
\end{split}
\label{eq: def current current pert to su(2)k}
\end{equation}
We take the coupling constant $g$ to be real.
The (unitary) field theory%
~(\ref{eq: def current current pert to su(2)k})
is often
 referred to as a non-Abelian Thirring (Gross-Neveu) model.
Suitable non-unitary generalizations
of the field theory%
~(\ref{eq: def current current pert to su(2)k})
compute 
(disorder average) moments of Green's functions
in a class of problems of Anderson localization in 
$d=2$ dimensions that we will investigate later on in this paper.

The one-loop beta function,
\begin{eqnarray}
\beta^{\ }_{g}=
\frac{{d}g}{{d}l}=
4 g^{2},
\end{eqnarray}
encodes the change in the coupling constant caused by 
the infinitesimal rescaling $\mathfrak{a}\to(1+{d}l)\mathfrak{a}$
of the short-distance cutoff $\mathfrak{a}$.
Thus, the current-current interaction
is (marginally) irrelevant (relevant) 
for $g<0$ ($g>0$)
with the free-fermion fixed point at $g=0$.

The $\mathrm{SU}(2)^{\ }_{k}$ Noether currents 
(\ref{eq: fermionic oscillator rep of su(2)})
are those appearing at
the non-trivial fixed point of the
Principal Chiral NL$\sigma$M on the SU(2) group manifold
with a Wess-Zumino term.%
~\cite{Novikov82}$^{-}$\cite{Polyakov83}
This has, the well-known (Euclidean) action
\begin{subequations}
\label{eq: SU(N) WZW action} 
\begin{equation}
S=
\frac{k}{16\pi}
\int d^2 x\,
\mathrm{tr}
\left(
\partial^{\ }_{\mu} G^{-1}\partial^{\ }_{\mu} G 
\right)
+
k\Gamma[G],
\end{equation}
where $G\in \mathrm{SU}(2)$
is a group element, and 
the integral $\Gamma[G]$
over a three-dimensional ball
$B$ with coordinates $r^{\ }_{\mu}$ and
whose boundary $\partial B$
is $d=2$-dimensional Euclidean space,
\begin{equation}
\Gamma[g]:= \frac{1}{24\pi}
 \int\limits_{B} d^{3}r\,
\epsilon^{\ }_{\mu\nu\lambda}
\mathrm{tr}\,
\left(
G^{-1}\partial^{\ }_{\mu}G\,
G^{-1}\partial^{\ }_{\nu}G\,
G^{-1}\partial^{\ }_{\lambda}G
\right)
\end{equation}
\end{subequations}
is the Wess-Zumino term.%
~\cite{Wess71}

The Noether currents which generate the 
SU(2)${\ }_{\mathrm{left}}$$\,\times\,$SU(2)${\ }_{\mathrm{right}}$
symmetry at the critical point of the WZW theory
can be fully represented by the fermionic expressions in 
Eq.~(\ref{eq: fermionic oscillator rep of su(2)}).
In the bosonic
(i.e., NL$\sigma$M)
representation, these currents are built out of 
first-order derivatives of the bosonic fields,
\begin{equation}
J^{\ }_{\alpha}\propto 
k\,
\mathrm{tr}
\left[
(\partial G)G^{-1}\sigma^{\ }_{\alpha}
\right],
\quad
\bar{J}^{\ }_{\alpha}\propto 
k\,
\mathrm{tr}
\left[
G^{-1}(\bar\partial G)\sigma^{\ }_{\alpha}
\right].
\label{eq: bosonized currents}
\end{equation}

The relationship~(\ref{eq: bosonized currents})
suggests that composite operators 
built out of monomials in the  currents%
~(\ref{eq: fermionic oscillator rep of su(2)})
in the WZW theory are the counterparts of
the high-gradient operators in NL$\sigma$M. For this reason,
we shall still call the former family of composite operators
high-gradient operators.
The ``classical'' counterparts of the 
high-gradient operators of order $s$
in the NL$\sigma$M are thus the homogeneous polynomials
\begin{eqnarray}
T^{    \alpha^{\ }_{1}\cdots    \alpha^{\ }_{s}
   \bar\alpha^{\ }_{1}\cdots\bar\alpha^{\ }_{s}}
J^{\ }_{\alpha^{\ }_{1}}
\cdots 
J^{\ }_{\alpha^{\ }_{s}}
\bar J^{\ }_{\bar \alpha^{\ }_{1}}
\cdots 
\bar J^{\ }_{\bar \alpha^{\ }_{s}}
\label{eq: classical hgo su(2)k}
\end{eqnarray}
of the left and right currents that are invariant under the diagonal
SU(2) symmetry group of the interacting theory.%
~\cite{footnote: invariant tensor}
The generating set of classical high-gradient operators of order $s$
is specified once all the linearly 
independent rank $2s$ tensors 
$
T^{\alpha\beta\cdots\gamma\delta\cdots}
$
in the adjoint representation of SU(2) 
that are invariant under
SU(2) transformations 
can be fully enumerated.
In turn, the most general 
SU(2) invariant tensor of even rank
in the adjoint representation
is the product of the Casimir tensor of rank 2.%
~\cite{Dittner72}

The high-gradient operators in Eq.~(\ref{eq: classical hgo su(2)k}) 
are classical in the sense that quantum fluctuations encoded 
through the Pauli principle (or, equivalently, 
through the underlying Dirac-Fermi sea)
in the free-fermion representation%
~(\ref{eq: fermionic oscillator rep of su(2)})
of the current algebra, have not yet been accounted for.
To account for these quantum fluctuations,
one needs to introduce a point-splitting 
procedure that allows for the proper normal ordering,
i.e., the correct subtraction of all short-distance singularities%
~\cite{DiFrancesco97}
\begin{equation}
\begin{split}
&
:
T^{    \alpha^{\ }_{1}\cdots\bar\alpha^{\ }_{s}}
     J^{\ }_{    \alpha^{\ }_{1}}
\cdots 
\bar J^{\ }_{\bar\alpha^{\ }_{s}}
:(z,\bar z)\equiv
\\
&
\qquad
\lim_{z^{\ }_{i}\to z}
\Big[
T^{    \alpha^{\ }_{1}\cdots\bar\alpha^{\ }_{s}}
     J^{\ }_{    \alpha^{\ }_{1}}(     z^{\ }_{1})
\cdots 
\bar J^{\ }_{\bar\alpha^{\ }_{s}}(\bar z^{\ }_{s})
\\
&
\qquad\qquad\quad
-\mbox{
(all short-distance singularities)
      }
\Big].
\end{split}
\label{eq: quantum hgo su(2)k}
\end{equation}

Two objects of the form~(\ref{eq: classical hgo su(2)k})
that are linearly independent classically
might not survive as a pair of distinct
quantum operators of the form%
~(\ref{eq: quantum hgo su(2)k})
after normal ordering has been implemented.
More precisely, one might anticipate that
the underlying free-fermion representation
of the current algebra must manifest itself 
as soon as the order $s$ 
becomes larger than the number $k$ of fermionic flavors 
by changing the book-keeping relating classical expressions labeled by
SU(2) tensors of rank $2s$ and quantum operators. 

Indeed, we are going to show that this is the mechanism that prevents
high-gradient operators of order $s>k$
from acquiring one-loop scaling dimensions smaller
than the smallest one-loop scaling dimensions associated with the set of
all high-gradient operators of order $s\leq k$.
In other words, the smallest
one-loop dimension associated with the set of \textit{all}
high-gradient operators is reached within the set of all
high-gradient operators of order $s\leq k$ when $g>0$. 
It is thus bounded from below when $g>0$.

Had we ignored the underlying free-fermion representation 
of the current algebra altogether,
we would have wrongly predicted that, when $g>0$, 
the one-loop dimensions associated with the
classical objects~(\ref{eq: classical hgo su(2)k})
are of a form similar to the ones in Eq.%
~(\ref{eq: anomalous scaling dimensions for O(N) high-gradient operators})
i.e.,
that the set of one-loop dimensions of high-gradient operators
is unbounded from below. On the other hand,
this classical prediction is recovered
in the limit $k\to\infty$ with $s/k\to0$. 
For this reason we shall separate the computation of
the most relevant one-loop dimension 
associated with high-gradient operators
of order $s$ into the case when $s\leq k$ and the case when $k< s$.

In this context, we would like to remind the reader that
the $\widehat{\mathrm{su}}(2)^{\ }_{k}$ WZW theories are known to
describe quantum critical points in the parameter space of 
quantum spin-$S$ antiferromagnetic chains when $k=2S$. 
Here, we observe that both, 
the number of relevant perturbations
and the number of independent local composite operators 
built out of the generators of SU(2) 
which are SU(2) singlets, grows with $S$. 
[For $S=1/2$ the algebra obeyed by the Pauli
matrices only allows one invariant SU(2) tensor of rank 2, 
the $2\times2$ unit matrix.]
On the other hand,
the strength of quantum fluctuations in
$\widehat{\mathrm{su}}(2)^{\ }_{k}$ WZW theories
decreases with increasing $k=2S$ 
for the same reason as the role of 
quantum fluctuations decreases with increasing
$S$ in quantum spin chains.

\subsection{
Anomalous dimensions of high-gradient operators
        }
\label{subsec: High-gradient operators when 1<s<k}

As the most general 
SU(2) invariant tensor of even rank
in the adjoint representation
is the product of the Casimir tensor of rank 2,\cite{Dittner72}
we define the three diagonal SU(2) invariants 
out of the three current bilinears
\begin{subequations}
\begin{equation}
\begin{split}
H
:=
C^{\alpha\beta} 
     J^{\ }_{\alpha}
\bar J^{\ }_{\beta},
\quad
A
:=
C^{\alpha\beta} 
J^{\ }_{\alpha}
J^{\ }_{\beta},
\quad
B
:=
C^{\alpha\beta} 
\bar J^{\ }_{\alpha}
\bar J^{\ }_{\beta},
\end{split}
\end{equation}
together with the 
SU(2) invariant
\begin{equation}
C:=AB.
\end{equation}
\end{subequations}
The space of the high-gradient operators
is then spanned by the family
\begin{eqnarray}
\left\{
H^s,
H^{s-2}C,\cdots,
H^{2} C^{[s/2]-1},
C^{[s/2]}
\right\}
\label{eq: def high gradiant order s if s<k}
\end{eqnarray}
made of $[s/2]+1$ ``classical'' operators.%
~\cite{footnote: choice HGO for s<k}
We call these operators ``classical'' because we have
not yet taken into account the short distance singularities
associated with the
definition of composite operators
(i.e., the ``Pauli principle'' discussed above).
As announced below Eq.~(\ref{eq: quantum hgo su(2)k}),
these singularities need to be subtracted
from the ``classical'' expressions%
~(\ref{eq: def high gradiant order s if s<k})
upon normal ordering.
We shall nevertheless ignore the issue of
normal ordering at first and compute the one-loop RG equation
for these ``un-regularized'' (or un-normal-ordered) operators,
a step of no consequence in the (``classical'') limit $s/k\to0$.
We shall then contrast this un-regularized calculation with 
the full quantum calculation for the special case of $k=1$,
i.e., when the proper normal ordering procedure has been accounted for.

We shall see that the calculation without normal ordering 
gives an infinite tower of high-gradient operators
that are all relevant to one-loop order
for sufficiently large $s$ and for $g>0$.
The one-loop spectrum of anomalous dimensions is identical to
the one in the $\mathrm{O}(N)/\mathrm{O}(N-1)$ NL$\sigma$M
when $N=3$. 
Indeed, once the normal ordering procedure is ignored, 
the high-gradient operators
(\ref{eq: def high gradiant order s if s<k})
are analogous to those discussed in Ref.\ \onlinecite{Castilla97},
and the calculations of the anomalous dimensions
in the $\hat{\mathrm{su}}(2)^{\ }_{k}$ WZW model
and
in the $\mathrm{O}(N)/\mathrm{O}(N-1)$ NL$\sigma$M
run along parallel tracks. 

The effect of normal ordering is weaker the smaller $s/k$ is, i.e., 
the closer proximity to the semi-classical limit
of the WZW theory.
To see this, consider the case when $k>s$.
The ``classical'' expression 
$
T^{    \alpha^{\ }_{1}\cdots\bar\alpha^{\ }_{s}}
     J^{\ }_{    \alpha^{\ }_{1}}
\cdots 
\bar J^{\ }_{\bar\alpha^{\ }_{s}}
(z,\bar z)$
for the composite operator made of a local product of
holomorphic and antiholomorphic currents
is modified upon normal ordering.
To leading order in a short-distance expansion, 
this classical expression is replaced by
\begin{eqnarray}
&& \!\!
 \sum_{
     \iota^{\ }_{1}\neq\cdots\neq    \iota^{\ }_{s}\neq
 \bar\iota^{\ }_{1}\neq\cdots\neq\bar\iota^{\ }_{s}=1
      }^{k}
 T^{    \alpha^{\ }_{1}\cdots    \alpha^{\ }_{s}
    \bar\alpha^{\ }_{1}\cdots\bar\alpha^{\ }_{s}}
 :\!
           J^{\ }_{\alpha^{\ }_{1}\iota^{\ }_{1}}
 \cdots 
           J^{\ }_{\alpha^{\ }_{s}\iota^{\ }_{s}}
 \!:\!(z)
\nonumber\\
&&\hspace*{27mm}{}\times
 :\!
      \bar J^{\ }_{\bar\alpha^{\ }_{1}\bar\iota^{\ }_{1}} 
 \cdots 
      \bar J^{\ }_{\bar\alpha^{\ }_{s}\bar\iota^{\ }_{s}} \!
 :\!(\bar{z})
 +\cdots.
\end{eqnarray}
Here, the terms included in the $\cdots$ arise 
from the OPE for the product 
$J^{\ }_{\alpha^{\ }_{i}\iota^{\ }_{i}}(z)
 J^{\ }_{\alpha^{\ }_{j}\iota^{\ }_{j}}(0)$
when any two flavor indices $\iota^{\ }_{i}$ and $\iota^{\ }_{j}$ 
are identical. Evidently, normal ordering (or the Pauli principle)
has a much more potent effect when $k<s$,
for the condition
$\iota^{\ }_{1}\neq\cdots\neq    \iota^{\ }_{s}\neq
 \bar\iota^{\ }_{1}\neq\cdots\neq\bar\iota^{\ }_{s}$
can then never be met so that the 
leading order term above is absent.
The operator contents with and without normal ordering
thus look very different.
When $k<s$, some operators in
the set (\ref{eq: def high gradiant order s if s<k})
completely disappear to leading order because of
Fermi statistics. This will be demonstrated explicitly for the case of $k=1$
[see Eqs.\ (\ref{eq: normal ordered JJ})
and (\ref{eq: normal ordered JJ second})
below],
for which we will show, after correctly taking into account
normal ordering, that all high-gradient operators 
which would be relevant classically (when $g > 0$)
disappear from the operator content.

\subsubsection{RG equation for un-regularized high-gradient operators}

To compute the \textit{leading} one-loop scaling dimensions
for the high-gradient operators%
~(\ref{eq: def high gradiant order s if s<k}),
we start from the field theory%
~(\ref{eq: def current current pert to su(2)k})
in which we substitute the action by
\begin{equation}
\begin{split}
S:=& \,
S^{\ }_{*}
+
g
\int\frac{d\bar zdz}{2\pi{i}}\,
\mathcal{O}^{\ }_{I}
\\
&
-
\sum_{
m,n=0
     }^{2m+n=s}
Z^{(s)}_{m,n}
\mathfrak{a}^{2s-2}
\int\frac{d\bar zdz}{2\pi{i}}\,
C^{m}H^n.
\end{split}
\label{eq: def pert action by g and Z's}
\end{equation}
To determine the one-loop dimensions
of the couplings 
$\{Z^{(s)}_{m,n}|2m+n=s\}$,
we do not need the full one-loop RG flows,
i.e., the RG equations for
 the coupling constants
up to and including order $Z^{(s)}_{m,n}Z^{(s)}_{p,q}$,
but
 only the linear in $Z^{(s)}_{m,n}$
contributions to the one-loop RG flows.
Thus, all we need are the OPE of
$
C^{m}H^{n}(z,\bar{z})$
with
$\mathcal{O}^{\ }_I(0)
$,
where the
 integers $m$ and $n$ satisfy $1<2m+n=s\leq k$.
Furthermore, we shall introduce the short-hand notation
\begin{equation}
\mathcal{A}
\times
\mathcal{B}=
\mathcal{C}
\Longleftrightarrow
\mathcal{A}(z,\bar z)
\mathcal{B}(0)=
\frac{1}{z\bar z}
\mathcal{C}(z,\bar z)
+
\cdots
\label{eq: def short hand OPE}
\end{equation}
for the OPE relating the operators 
$\mathcal{A}$,
$\mathcal{B}$,
and
$\mathcal{C}$.
Here,
the dots are meant to contain not only regular terms of
zeroth and higher order
in $z$ or $\bar z$ but also second and higher order poles in $z$ or $\bar z$.

As an intermediary step, one verifies that the
OPE (\ref{eq: def short hand OPE}) 
between the building blocks $H$ and $C$ 
with $\mathcal{O}^{\ }_{I}$
(observe that $\mathcal{O}^{\ }_{I}=H$)
are 
\begin{eqnarray}
\Wick{7mm}
\Wickunder{7mm}H
{\times}
\mathcal{O}^{\ }_{I}
&=&
-4
H,
\qquad
\Wick{7mm}
\Wickunder{7mm} 
C{\times}
\mathcal{O}^{\ }_{I}=
0.
\label{eq: Step 1}
\end{eqnarray}
Here, we introduced yet another short-hand notation
$\Wick{8mm}\mathcal{A} \cdots \mathcal{B}$
or
$\Wickunder{8mm}\mathcal{A} \cdots \mathcal{B}$,
by which we mean that one current in $\mathcal{A}$
and one current in  $\mathcal{B}$ are contracted 
with the rule 
\begin{equation}
\begin{split}
&
J^{\ }_{\alpha\iota}(z) 
J^{\ }_{\beta \iota'}(0)=
\delta^{\ }_{\iota\iota'}\!\!
\left(
\frac{C^{\ }_{\alpha\beta}}{z^{2}}
\!+\!
\frac{{i}}{z}
 f^{\ }_{\alpha\beta}{}^{\gamma}
J^{\ }_{\gamma\iota}(0)
\!+\!
\cdots
\right),
\\
&
\bar J^{\ }_{\alpha\iota }(\bar z) 
\bar J^{\ }_{\beta \iota'}(    0)=
\delta^{\ }_{\iota\iota'}\!\!
\left(
\frac{C^{\ }_{\alpha\beta} }{\bar z^{2}}
\!+\!
\frac{{i}}{\bar z}
 f^{\ }_{\alpha\beta}{}^{\gamma}
\bar J^{\ }_{\gamma\iota}(0)
\!+\!
\cdots
\right),
\\
&
     J^{\ }_{\alpha\iota }(z) 
\bar J^{\ }_{\beta \iota'}(0)=
0,
\end{split}
\label{eq: OPE for the currents for each species}
\end{equation}
for any
$
\alpha,\beta=1,2,3
$
and
$
\iota,\iota'=1,\cdots,k
$
at the free-fermion fixed point $g=0$, where
\begin{equation}
\begin{split}
 J^{\ }_{\alpha\iota}:=
 \psi^{a \dag}_{\ \iota}
 \frac{\left(\sigma^{\ }_{\alpha}\right)^{\ }_{a}{}^{b}}{2}
 \psi^{\  }_{b \iota},
 \quad
 \bar J^{\ }_{\alpha\iota}:=
 \bar \psi^{a \dag}_{\ \iota}
 \frac{\left(\sigma^{\ }_{\alpha}\right)^{\ }_{a}{}^{b}}{2}
 \bar \psi^{\  }_{b \iota}.
 \end{split}
\label{eq: flavor currents}
\end{equation}
When $\mathcal{A}$ and $\mathcal{B}$ consist of
more than one $J^{\ }_{\alpha}$ or $\bar{J}^{\ }_{\alpha}$,
and when there are many possible Wick contractions 
between $\mathcal{A}$ and $\mathcal{B}$, 
the short-hand notations
$\Wick{8mm}\mathcal{A} \cdots \mathcal{B}$,
$\Wickunder{8mm}\mathcal{A} \cdots \mathcal{B}$
and
$\Wick{8mm}\Wickunder{8mm}\mathcal{A} \cdots \mathcal{B}$
mean the resulting operator 
obtained by taking all possible such Wick contractions.
One also verifies that the
OPE (\ref{eq: def short hand OPE}) 
between the building blocks $HH$, $CH$, and $CC$ 
with $\mathcal{O}^{\ }_{I}$
are 
\begin{equation}
\begin{split}
\Wick{10mm}H 
\Wickunder{7mm}H 
{\times}
\mathcal{O}^{\ }_I
&=
 -4 H^{2}
+ 4 C,
\\
\Wick{10mm}
C
\Wickunder{7mm}
H
{\times}
\mathcal{O}^{\ }_I
&=
\Wick{10mm}
C
\Wickunder{7mm}
C
{\times} 
\mathcal{O}^{\ }_I
=
0.
\end{split}
\label{eq: Step 2}
\end{equation}
We then infer that, for any pair $(m,n)$ of 
positive integer
that satisfies $1<2m+n=s\leq k$,
\begin{equation}
\begin{split}
C^{m}H^{n}
{\times} 
\mathcal{O}^{\ }_{I}
=&
\hphantom{+}
\Wick{7mm}\Wickunder{7mm}C 
{\times} 
\mathcal{O}^{\ }_I
\times 
m C^{m-1}H^{n}
\\
&
+
\Wick{7mm}\Wickunder{7mm}H
{\times} 
\mathcal{O}^{\ }_I
\times 
n C^{m}H^{n-1}
\\
&
+
\Wick{10mm}C\Wickunder{7mm} H
{\times} 
\mathcal{O}^{\ }_{I}
\times
mn C^{m-1}H^{n-1}
\\
&
+
\Wick{10mm}C\Wickunder{7mm} C
{\times} 
\mathcal{O}^{\ }_{I}
\times
\frac{m(m-1)}{2} C^{m-2}H^{n}
\\
&
+
\Wick{10mm}H\Wickunder{7mm} H 
{\times} 
\mathcal{O}^{\ }_{I}
\times
\frac{n(n-1)}{2} C^{m}H^{n-2}
\\
=&
-2
n\left(n+1\right)
C^{m} H^{n}
\\
&
+
2
n(n-1) 
C^{m+1}H^{n-2}.
\end{split}
\label{eq: OPE's between H and CmHN}
\end{equation}
The contributions to the RG equations obeyed by the couplings
$Z^{(s)}_{m,n}$ where $1<2m+n=s\leq k$
needed to extract the spectrum of one-loop dimensions are
\begin{equation}
\begin{split}
\frac{{d}Z^{(s)}_{m,n}}{{d}l}=& \,
\big(
2
-
2s
\big)
Z^{(s)}_{m,n}
+
4gn(n+1)
Z^{(s)}_{m,n}
\\
&
-
4g  
(n+2)(n+1)
Z^{(s)}_{m-1,n+2}
+
\cdots.
\end{split}
\label{eq: linearized RG flows for Z(m,n)}
\end{equation}
Here, the dots include non-linear contributions of
second order in $g$ or $Z^{(s)}_{m,n}$.

The linearized RG flows
(\ref{eq: linearized RG flows for Z(m,n)})
are closed. This is a justification a posteriori 
for neglecting the RG effects of current monomials with
repeating flavor indices.
The linearized RG flows
(\ref{eq: linearized RG flows for Z(m,n)})
have a lower triangular structure,
i.e., there is no feedback effect on the flow of
a high-gradient operator of order $s$ 
from lower-order high-gradient operators.
Thus,  we conclude that the leading 
$[s/2]+1$ one-loop scaling dimensions associated
with the family of high-gradient operators%
~({\ref{eq: def high gradiant order s if s<k})
when $k\geq s=2m+n$ are given by
\begin{eqnarray}
x^{(s)}_{m,n}&=&
2(2m+n)
-
4 g n(n+1).
\end{eqnarray}
Observe that the spectrum of
anomalous dimensions%
~\cite{footnote: our conventions for dimensions}
\begin{equation}
\gamma^{(s)}_{m,n}:=
-
4 g n(n+1),
\qquad
2m+n=s
\label{eq: gamma(s)m,n is one sided for su(2) level k<s}
\end{equation}
is one sided with respect to 0.
When $g\leq0$ these anomalous dimensions are positive, i.e.,
the scaling dimensions are larger than their engineering value.
The opposite happens when $g\geq0$, i.e., when the current-current
perturbation is (marginally) relevant.
 When $g>0$ and for a given $1<s\leq k$,
the smallest one-loop anomalous dimension
occurs for the pair $(m,n)=(0,s)$,
\begin{eqnarray}
\gamma^{(s)}_{\mathrm{min}}&:=&
\min_{2m+n=s}
\gamma^{(s)}_{m,n}
=
-
4gs(s+1).
\label{eq: max scaling dimension when g>0}
\end{eqnarray}
For $g>0$, 
the quadratic dependence on $s$ can overcome the linear dependence on $s$
in the one-loop dimension
$x^{(s)}_{\mathrm{min}}:=2s+\gamma^{(s)}_{\mathrm{min}}$.
If the order $s$ ($1<s\leq k$) is allowed to be sufficiently large,
the one-loop dimension $x^{(s)}_{\mathrm{min}}$
decreases past the value 2 and eventually becomes negative.
The quadratic dependence on $s$ is reminiscent of that for the
one-loop dimensions%
~(\ref{eq: anomalous scaling dimensions for O(N) high-gradient operators})
in the $\mathrm{O}(N)/\mathrm{O}(N-1)$ NL$\sigma$M.
However, in contrast to the 
$(2+\epsilon)$-dimensional $\mathrm{O}(N)/\mathrm{O}(N-1)$ NL$\sigma$M
at its non-trivial fixed point $t^{*}$,
a value smaller than 2 for the one-loop dimensions
$x^{(s)}_{\mathrm{min}}$
is not a threat to the internal stability of the WZW fixed point $g=0$
since it occurs along a flow to strong coupling.
Moreover, it is known that in $d=2$ dimensions the 
$\mathrm{O}(3)/\mathrm{O}(2)$ NL$\sigma$M
with theta term at $\theta =\pi$  flows
in the infrared into
the level $k=1$  SU(2) WZW fixed point.
While the spectrum of one-loop dimensions
of high-gradient operators at the WZW fixed point is
bounded from below (as we will recall below),
the spectrum of these operators
is unbounded from below in the
weakly coupled $2$-dimensional
$\mathrm{O}(3)/\mathrm{O}(2)$ NL$\sigma$M
(the presence of the theta term does not affect this result).

\subsubsection{
Normal ordering revisited
              }
\label{subsec: high-gradient operators when s>k }

We shall illustrate the effects of the Fermi statistics
for the family of high-gradient operators%
~({\ref{eq: def high gradiant order s if s<k})
when $s=2$ for the case of a (marginally) relevant ($g>0$) 
current-current interaction.
We shall then show for the special case of $k=1$ and $s=2$
that the two one-loop dimensions associated with 
the family of high-gradient operators%
~({\ref{eq: def high gradiant order s if s<k})
are unchanged, to one loop order, i.e.,
\begin{equation}
x^{(s)}_{0,2}=
x^{(s)}_{1,0}=4.
\end{equation}

We start from the family of high-gradient operators%
~({\ref{eq: def high gradiant order s if s<k})
with $s=2$. For clarity of presentation,
we rename the two members
of this family,
\begin{equation}
\mathcal{O}^{\ }_{1}\equiv
C^{\alpha\beta}J^{\ }_{\alpha}J^{\ }_{\beta}
C^{\gamma\delta}\bar J^{\ }_{\gamma}\bar J^{\ }_{\delta},
\quad
\mathcal{O}^{\ }_{2}\equiv
C^{\alpha\beta}J^{\ }_{\alpha}\bar J^{\ }_{\beta}
C^{\gamma\delta}J^{\ }_{\gamma}\bar J^{\ }_{\delta}.
\label{eq: classical O1 and O2 HGO}
\end{equation}
As implied by Eq.~(\ref{eq: quantum hgo su(2)k})
these are two classical expressions.
The two quantum expressions involve 
point splitting and normal ordering
as in Eq.\ (\ref{eq: quantum hgo su(2)k}).

Without loss of generality, we consider only the left current sector.
Normal ordering of
\begin{equation}
\begin{split}
J^{\ }_{\alpha}(z)
J^{\ }_{\beta }(0)=&
\frac{k C^{\ }_{\alpha\beta}}{z^{2}}
+
\frac{{i}}{z}
\epsilon^{\ }_{\alpha\beta\gamma}J^{\ }_{\gamma}(0)
+
\frac{{i}}{2}
\epsilon^{\ }_{\alpha\beta\gamma}
\partial 
J^{\ }_{\gamma}
(0)
\\
&+
\frac{\delta^{\ }_{\alpha\beta}}{4}
\sum_{\iota=1}^{k}
:
\left(
\psi^{a\dag}_{\ \iota}
\partial 
\psi^{\ }_{a\iota}
-
\partial 
\psi^{a\dag}_{\ \iota}
\psi^{\ }_{a\iota}
\right)
:
(0)
\\
&
+
\sum_{\iota,\iota'=1}^{k}
:
\psi^{a\dag}_{\ \iota}
\frac{(\sigma^{\ }_{\alpha})^{\ }_{a}{}^{b}}{2}
\psi^{\ }_{b\iota}
\psi^{c\dag}_{\ \iota'}
\frac{(\sigma^{\ }_{\beta})^{\ }_{c}{}^{d}}{2}
\psi^{\ }_{d\iota'}
:(0)
\\
&
+\cdots
\end{split}
\label{eq: OPE two currents made of fermions}
\end{equation}
amounts to the subtraction
from  Eq.~(\ref{eq: OPE two currents made of fermions})
of the terms singular in the limit $z\to0$,
\begin{equation}
\begin{split}
\label{eq: normal ordered JJ}
&
:J^{\ }_{\alpha}J^{\ }_{\beta}:(0) = 
\sum_{\iota\neq\iota'=1}^{k} 
J^{\ }_{\alpha \iota}
J^{\ }_{\beta \iota'}
(0)
+
\frac{{i}}{2}
\epsilon^{\ }_{\alpha\beta\gamma}
\partial J^{\ }_{\gamma}(0)
\\
&
+
\frac{\delta_{\alpha\beta}}{4}\sum_{\iota=1}^{k}
:
\big(
\partial
\psi^{a\dag}_{\ \iota}
\psi^{\ }_{a \iota}
-
\psi^{a\dag}_{\ \iota}
\partial 
\psi^{\ }_{a \iota}
-
\psi^{a\dag}_{\ \iota}
\psi^{\ }_{a \iota}
\psi^{b\dag}_{\ \iota}
\psi^{\ }_{b \iota}
\big)
:(0)
\end{split}
\end{equation}
for $\alpha,\beta=1,2,3$.
The proper quantum interpretation of the
classical currents~(\ref{eq: classical O1 and O2 HGO})
is then
\begin{equation}
\begin{split}
&
\label{eq: normal ordered JJ second}
:\mathcal{O}^{\ }_{1}:(z,\bar z)=
4
\sum_{\alpha,\beta=1}^{3}
:     J^{\ }_{\alpha}     J^{\ }_{\alpha}:(z)
:\bar J^{\ }_{\beta }\bar J^{\ }_{\beta }:(\bar z),
\\
&
:\mathcal{O}^{\ }_{2}:(z,\bar z)=
4\sum_{\alpha,\beta=1}^{3}
:     J^{\ }_{\alpha}     J^{\ }_{\beta}:(z)
:\bar J^{\ }_{\alpha}\bar J^{\ }_{\beta}:(\bar z).
\end{split}
\end{equation}

\subsubsection{
High-gradient operators when $k=1$
              }

When $k=1$,
the summation over unequal flavors disappears
in Eq.~(\ref{eq: normal ordered JJ}).
(Observe in passing that
$:\!\mathcal{O}^{\ }_{1}\!:$
is then proportional to one component of the energy-momentum
stress tensor.)
One then verifies the OPE
\label{eq: correct OPE}
\begin{equation}
\begin{split}
&
:\!\mathcal{O}^{\ }_{1}\!:
\times \mathcal{O}^{\ }_{I}=
3:\!\mathcal{O}^{\ }_{1}\!:
- \,
9
:\!\mathcal{O}^{\ }_{2}\!: \,
,
\\
&
:\!\mathcal{O}^{\ }_{2}\!:
\times
\mathcal{O}^{\ }_{I}= \,
:\!\mathcal{O}^{\ }_{1}\!:
- \,
3:\!\mathcal{O}^{\ }_{2}\!:.
\end{split}
\end{equation}
If we diagonalize the linearized one-loop RG flows for 
the coupling $Z^{(2)}_{1,0}$ 
associated with 
$:\!\mathcal{O}^{\ }_{1}\!:$
and
the coupling $Z^{(2)}_{0,2}$ 
associated with 
 $:\!\mathcal{O}^{\ }_{2}\!:$,
we find that their one-loop dimensions 
remain equal to their engineering dimensions,
\begin{equation}
x^{(2)}_{1,0}=
x^{(2)}_{0,2}=
4.
\end{equation}

The lesson that we draw from the example $s=2$ and $k=1$
is that it is necessary to use normal ordering 
to properly define composite operators.
Had we not used normal ordering, we would have incorrectly 
predicted that there are infinitely many high-gradient operators 
which become relevant, at one-loop order, for large
enough $s$ and for $g>0$.
We believe that for a generic value of $k$, 
there is no infinity of
one-loop relevant high-gradient operators.
Only a finite number of high-gradient operators become
relevant, at one-loop order,
for large enough $s$ and for $g>0$ when $k>1$.

In the next section, we turn attention 
to a  non-unitary WZW model of relevance to the problem
of Anderson localization to investigate whether
the loss of unitarity opens the door to an infinity
of relevant high-gradient operators.

\section{ 
High-gradient operators and $\widehat{\mathrm{gl}}(M|M)_{k}$ WZW theories
        }
\label{sec: gl(M|N) level k=1}

An interesting example of a problem of Anderson localization
in two dimensions which possesses a special so-called sublattice 
(or chiral) symmetry (SLS) and TRS (thus belonging to
the ``chiral-orthogonal'' symmetry class BDI
in the classification scheme of Zirnbauer, and Altland and 
Zirnbauer%
~\cite{Zirnbauer96}$^{-}$\cite{Heinzner05})
is as follows.  
Consider a tight-binding model of fermions on a honeycomb lattice 
with random real-valued hopping matrix elements of non-vanishing mean,
which do not connect the same sublattice
(so that SLS is preserved).
[A related  realization of the same problem of Anderson localization 
is provided by a random tight-binding model on a square lattice with flux-$\pi$
through every plaquette, introduced in Ref.~\onlinecite{Hatsugai97}.]
In the absence of disorder this band structure is known to exhibit the 
energy-momentum dispersion law of two species of (relativistic) Dirac
fermions at two points in the Brillouin zone at low energy
near the Fermi level (at zero energy).
It was shown in Ref. \onlinecite{Guruswamy00} that the 
SLS-preserving disorder discussed above 
leads to a theory for the disorder averages which, in the
supersymmetric formulation,\cite{Efetov97}
is a ${\mathrm{GL}}(2N|2N)$ Thirring (Gross-Neveu) model.
In other words, the problem of two-dimensional 
Anderson localization
on the honeycomb lattice preserving SLS and TRS, is described by a set 
of Dirac fermions (and SUSY boson partners) perturbed by a 
current-current interaction of 
the Noether currents
of its underlying  ${\mathrm{GL}}(2N|2N)$ (super) symmetry.
The interaction strength corresponds to the strength of the disorder.
The system of free Dirac fermions (and SUSY boson partners) 
is well known~\cite{Witten84,Bocquet00}
to be described by a WZW model on the supergroup
${\mathrm{GL}}(2N|2N)$ 
with $\widehat{\mathrm{gl}}(2N|2N)^{\ }_{k}$ 
conformal Kac-Moody current algebra symmetry at level $k=1$.

This section is devoted to the one-loop RG analysis of
high-gradient operators
in the perturbed $\widehat{\mathrm{gl}}(M|M)_{k}$ WZW theory.
The main result of this section and of this article
applies to the case of level $k=1$ of relevance
for the random tight-binding models discussed above.
In order to state this result, 
we first need to 
recall from Ref. \onlinecite{Guruswamy00} that 
the  ${\mathrm{GL}}(2N|2N)$ Thirring (Gross-Neveu) models 
possess two coupling constants;
one, $g^{\ }_{\mathrm{M}}$, 
which does not flow under the renormalization group (RG) and another, 
$g^{\ }_{\mathrm{A}}$, which flows logarithmically under the RG
and a rate dependent on $g^{\ }_{\mathrm{M}}$. 
Our main result then suggests that
all higher-order gradient operators are more irrelevant
in the presence of the current-current interaction
with $g^{\ }_{\mathrm{M}}>0$ 
than at zero coupling $g^{\ }_{\mathrm{M}}=0$. 
A positive $g^{\ }_{\mathrm{M}}$
can be interpreted~\cite{Guruswamy00}
as the variance of the disorder strength
in the random tight-binding model in symmetry class BDI.
For the opposite sign of the coupling constant 
$g^{\ }_{\mathrm{M}}<0$,
on the other hand,
higher-order gradient operators have a spectrum of one-loop dimensions
that is unbounded from below very much as in Eq.%
~(\ref{eq: anomalous scaling dimensions for O(N) high-gradient operators}).
In the context of Anderson localization,
the case with $g^{\ }_{\mathrm{M}}<0$
describes the  surface state of a
three-dimensional topological insulator
in the chiral-symplectic class (symmetry class CII)
of Anderson localization.%
\cite{Schnyder08,Ryu09}

As in Sec.~\ref{sec: su2},
we are going to distinguish two limits. 
In the first (classical) limit,
\begin{equation}
M\to\infty,
\qquad
k\to\infty,
\label{eq: extreme classical limit}
\end{equation}
OPEs between the high-gradient operators can be obtained without 
any reference to the composite nature of the currents. 
One then recovers a spectrum of
one-loop scaling dimensions 
for high-gradient operators that mimics closely
that of the NL$\sigma$Ms discussed above.
The second limit,
\begin{equation}
M=1,2,3,\cdots,
\qquad
k=1,
\label{eq: extreme quantum limit}
\end{equation}
is the opposite extreme to the first one in that the 
normal ordering of the currents and thus of the high-gradient operators 
is essential and changes dramatically the spectrum of 
one-loop scaling dimensions from the ``classical'' limit%
~(\ref{eq: extreme classical limit}).

\subsection{
Definitions
           }

Our starting point is a two-dimensional conformal
field theory characterized by the current
algebra~\cite{Guruswamy00}
\begin{subequations}
\label{eq: def gl(M,N) current algebra of level k}
\begin{equation}
\begin{split}
J^{\,B}_{A}{ }(z)
J^{\,D}_{C}(0)=&
\frac{k\mathfrak{c}^{BD}_{AC}}{z^{2}}
+
 \frac{1}{z}
\left[
\mathfrak{d}^{B}_{C}
J^{\,D}_{A}(0)
+
\mathfrak{e}^{BD}_{AC}
J^{\,B}_{C}(0)
\right]
\\
&
+
\cdots,
\\
\bar J^{\,B}_A(\bar z)
\bar J^{\,D}_C(0)=&
\frac{k\mathfrak{c}^{BD}_{AC}}{\bar z^{2}}
+
\frac{1}{\bar z}
\left[
\mathfrak{d}^{B}_{C}
\bar J^{\,D}_{A}(0)
+
\mathfrak{e}^{BD}_{AC}
\bar J^{\,B}_{C}(0)
\right]
\\
&
+
\cdots,
\\
     J^{\,B}_A(z)
\bar J^{\,D}_C(0)=&
0,
\end{split}
\end{equation}
where
\begin{equation}
\begin{split}
\mathfrak{c}^{BD}_{AC}:=
(-)^{B+1}
\delta^{D}_{A}
\delta^{B}_{C},
\end{split}
\end{equation}
and
\begin{equation}
\mathfrak{d}^{B}_{C}=
-
(-)^{BC}\delta^{B}_{C},
\quad
\mathfrak{e}^{BD}_{AC}=
(-)^{BC+D(B+C)}\delta^{D}_{A},
\label{eq: structure constants gl(M|N)}
\end{equation} 
\end{subequations}
with the indices $A,B,C,D=1,\cdots,M+N$,
where 
$\delta^{B}_{C}$ denotes the Kronecker delta.
The capitalized indices 
$A$,
$B$,
$C$,
and
$D$
also carry a grade which is either 0 for $M$ out of the $M+N$
values that they take or 1 for the remaining $N$ values.
It is the grade of the indices $A$ and $B$ that enters 
expressions such as $(-)^{A}$ or $(-)^{AB}$.
The grade $0$ ($1$) will shortly be associated with
bosons (fermions).
The positive integer $k$ is the level of the current algebra%
~(\ref{eq: def gl(M,N) current algebra of level k}).
The current algebra%
~(\ref{eq: def gl(M,N) current algebra of level k})
is associated with the Lie superalgebra
$\mathrm{gl}(M|N)$
defined by the structure constants
Eq.\ (\ref{eq: structure constants gl(M|N)})
for $A,B,C,D=1,\cdots,M+N$.
When $N=0$, the structure constants%
~(\ref{eq: structure constants gl(M|N)})
reduce to
\begin{equation}
\mathfrak{d}^{B}_{C}=
-
\delta^{B}_{C},
\qquad
\mathfrak{e}^{BD}_{AC}=
+\delta^{D}_{A},
\label{eq: structure constants gl(M|0)}
\end{equation}
for $A,B,C,D=1,\cdots,M$. These define
the Lie algebra gl($M$) of the non-compact Lie group GL$($M$)$.
When $M=0$, the structure constants%
~(\ref{eq: structure constants gl(M|N)})
reduce to
\begin{equation}
\mathfrak{d}^{B}_{C}=
+\delta^{B}_{C},
\qquad
\mathfrak{e}^{BD}_{AC}=
-\delta^{D}_{A},
\label{eq: structure constants gl(0|N)}
\end{equation}
for $A,B,C,D=1,\cdots,N$. These define
the Lie algebra u($N$) of the compact Lie group U($N$).

There exists a free-fermion and free-boson realization of
the current algebra%
~(\ref{eq: def gl(M,N) current algebra of level k})
defined by the action%
~\cite{footnote: Einstein convention}
\begin{subequations}
\label{eq: free fermion and free boson rep gl(M|N) level k}
\begin{equation}
S^{\ }_{*}:=
\sum_{\iota=1}^{k}
\int \frac{d\bar zd z}{2\pi{i}}
\left(
    \psi^{A\dag}_{\ \ \iota}
\bar \partial\,
    \psi^{\ }_{A\iota}
+
\bar \psi^{A\dag}_{\ \ \iota}
     \partial\,
\bar \psi^{\ }_{A\iota}
\right) 
\label{eq: def S* gl(M|N)}
\end{equation}
with the partition function
\begin{eqnarray}
Z^{\ }_{*}:=
\int\mathcal{D}[\psi^{\dag},\psi,
\bar{\psi}^{\dag},\bar{\psi}
]
\exp(-S^{\ }_{*}),
\end{eqnarray}
\end{subequations}
where it is understood that
$\psi^{\ }_{A\iota}$ and $\bar\psi^{\ }_{A\iota}$ are
complex-valued integration variables
for the $M$ values of $A$ with grade 0
while $\psi^{\ }_{A\iota}$ and $\bar\psi^{\ }_{A\iota}$
are Grassmann-valued integration variables
for the $N$ values of $A$ with grade 1, regardless
of the value taken by the flavor index $\iota=1,\cdots,k$.
The current algebra (\ref{eq: def gl(M,N) current algebra of level k})
is then realized by the representation
\begin{equation}
J^{\,B}_{A}:=
\sum_{\iota=1}^{k}
:\!
\psi^{\ }_{A\iota}
\psi^{B\dag}_{\ \ \iota}
\!:\,
,
\qquad
\bar J^{\,B}_{A}:=
\sum_{\iota=1}^{k}
:\!
\bar \psi^{\ }_{A\iota}
\bar \psi^{B\dag}_{\ \ \iota}
\!:,
\label{eq: def GL(M|N) currents as free spinors}
\end{equation}
as follows from the OPE
\begin{equation}
\begin{split}
&
\psi^{\ }_{A\iota}(z)
\psi^{B\dag}_{\iota'}(0)=
(-1)^{AB+1}
\psi^{B\dag}_{\iota'}(z)
\psi^{\ }_{A\iota}(0)=
\frac{\delta^{\ }_{\iota\iota'}\delta^{B}_{A}}{z},
\\
&
\bar{\psi}^{\ }_{A\iota}(\bar{z}) 
\bar{\psi}^{B\dag}_{\iota'}(0)=
(-1)^{AB+1}
\bar{\psi}^{B\dag}_{\iota'}(\bar{z}) 
\bar{\psi}^{\ }_{A\iota}(0)=
\frac{\delta^{\ }_{\iota\iota'}\delta^{B}_{A}}{\bar z},
\\
&
\psi^{\ }_{A   \iota }(z) 
\bar\psi^{B\dag}_{\ \ \iota'}(     0)=0,
\end{split}
\end{equation}
with $A,B=1,\cdots,M+N$ and $\iota=1,\cdots,k$.

The expressions in 
Eq.~(\ref{eq: def GL(M|N) currents as free spinors})
form a representation
of the $\widehat{\mathrm{gl}}(M|N)^{\ }_{k}$ 
current algebra in terms of free fermions.
There are two Casimir invariants of rank 2 in $\mathrm{gl}(M|N)$
that we use to perturb the free field theory 
(\ref{eq: free fermion and free boson rep gl(M|N) level k})
with two types of current-current interactions,
both of which are invariant under the global $\mathrm{GL}(M|N)$
symmetry,\cite{Guruswamy00}
\begin{equation}
\begin{split}
&
Z:=
\int\mathcal{D}[\psi^{\dag},\psi,
\bar{\psi}^\dag,\bar{\psi}]
\exp(-S),
\\
&
S:=
S^{\ }_{*}
+
\int \frac{d\bar zdz}{2\pi{i}}
\left(
\frac{g^{\ }_{\mathrm{A}}}{2\pi} 
\mathcal{O}^{\ }_{\mathrm{A}}
+
\frac{g^{\ }_{\mathrm{M}}}{2\pi}
\mathcal{O}^{\ }_{\mathrm{M}}
\right),
\\
&
\mathcal{O}^{\ }_{\mathrm{A}}:=
- 
     J^{\,A}_{A}\,
(-1)^A\, 
\bar J^{\,B}_{B}\,
(-1)^B
\equiv
-
\mathrm{str}\,     J\, 
\mathrm{str}\,\bar J,
\\
& 
\mathcal{O}^{\ }_{\mathrm{M}}:=
-
     J^{\,B}_{A}
\bar J^{\,A}_{B}
(-1)^A
\equiv
 -
\mathrm{str}\,
\left(
J\bar J
\right).
\label{eq: glMN perturbed by two current-current int.}
\end{split}
\end{equation}
Formally, one may allow the coupling constants
$g^{\ }_{\mathrm{A}}$
and
$g^{\ }_{\mathrm{M}}$
to take on any real (i.e., positive or \textit{negative})
values. However, to make connection with the 
above mentioned two-dimensional 
tight-binding models
in symmetry class BDI
of Anderson localization,
we must demand that
$g^{\ }_{\mathrm{A}}$
and
$g^{\ }_{\mathrm{M}}$
be positive.
(See Appendix~\ref{sec: HWK model}.)

The ``classical'' counterparts to the high-gradient operators of order $s$
in Eq.~(\ref{eq: classical hgo su(2)k})
are the homogeneous polynomials
\begin{equation}
T^{\,A^{\ }_{1}\cdots\,A^{\ }_{s}\,\bar A^{\ }_{1}  \cdots\,\bar A^{\ }_{s}}
 _{B^{\ }_{1}\,\cdots  B^{\ }_{s}  \bar B^{\ }_{1}\,\cdots  \bar B^{\ }_{s}}
J^{\,B^{\ }_{1}}_{A^{\ }_{1}}
\cdots 
J^{\,B^{\ }_{s}}_{A^{\ }_{s}}
\bar J^{\,\bar B^{\ }_{1}}_{\bar A^{\ }_{1}}
\cdots 
\bar J^{\,\bar B^{\ }_{s}}_{\bar A^{\ }_{s}}
\label{eq: classical hgo gl(M|N)k}
\end{equation}
of the left and right currents that are invariant under the diagonal
GL($M|N$) symmetry group of the interacting theory.%
~\cite{footnote: invariant tensor}
The set of ``classical'' high-gradient operators
of order $s$ is specified once all the linearly independent 
rank $2s$ invariant tensors
$
T^{\,A^{\ }_{1}\cdots\,A^{\ }_{s}\,\bar A^{\ }_{1} \cdots\,\bar A^{\ }_{s}}
 _{B^{\ }_{1}\,\cdots  B^{\ }_{s}  \bar B^{\ }_{1}\,\cdots \bar B^{\ }_{s}}
$
in the adjoint representation of GL($M|N$)
which are invariant under GL($M|N$) transformations
have been enumerated. 
At the quantum level, normal ordering
defines the quantum high-gradient operators of order $s$
as in Eq.\ (\ref{eq: quantum hgo su(2)k}).

We are now going to specialize to the case $M=N$
where the beta function for the coupling constant
$g^{\ }_{\mathrm{M}}$ vanishes identically, 
an exact result.~\cite{Guruswamy00}
(As already mentioned, the other coupling constant
$g^{\ }_{\mathrm{A}}$  flows logarithmically
at a rate set by 
$g^{\ }_{\mathrm{M}}$.)
The sector which we loosely denote by
\begin{equation}
\mathrm{PSL}(M|M)\sim 
\mathrm{GL}(M|M)/\mathrm{U}(1)\times
\mathrm{U}(1)
\label{eq: PSL and GL}
\end{equation} 
remains scale (conformally) invariant
for any value of $g^{\ }_{\mathrm{M}}$.
More specifically, $\mathrm{PSL}(M|M)$ is obtained 
by first factoring out the U(1) subgroup
thereby obtaining the subgroup 
$\mathrm{SL}(M|M)$ of $\mathrm{GL}(M|M)$,
followed in a second step by the ``gauging away''
of the states carrying the $\mathrm{U}(1)$ 
charges under
$
j:= 
J^{\ A}_{A}
$
and
$
\bar{j}:= 
\bar{J}^{\ A}_{A}
$.%
~\cite{Guruswamy00,Bershadsky99,Berkovits99}
This turns out to realize a line of 
RG fixed points (and conformal field theories)
labeled by the coupling constant $g^{\ }_{\mathrm{M}}$.%
~\cite{Guruswamy00}

\subsection{
High-gradient operators
when $M,k\to\infty$
           }
\label{subsec: largest HGO order s in GL(N|N) at N=k=infty}

We are going to show that, when $k$ and $M$ are very large, the spectrum
for the one-loop scaling dimensions of high-gradient operators
shares the same structure as that in Eq.%
~(\ref{eq: anomalous scaling dimensions for O(N) high-gradient operators}).
It will become clear by comparison to the case of $k=1$ that
the limit  $M,k\to\infty$ is the extreme ``classical'' limit
whereas the limit $k=1$ is the extreme ``quantum'' limit.

We restrict the family of ``classical'' high-gradient operators to
objects of the form
\begin{eqnarray}
\mathrm{str}\,
\left(
J\bar{J}JJ\bar{J}J 
\right) 
\mathrm{str}\,
\left(
J\bar{J}\bar{J} 
\right)
  \cdots,
\label{eq: def HGO if M,k to infty}
\end{eqnarray}
i.e., to diagonal GL($M|M$)-invariant
monomials of order $s$ in both the holomorphic and antiholomorphic
currents. For any given order $s$, 
the engineering dimensions are all equal and given by $2s$.
This degeneracy is lifted to first order in the coupling constant
$g^{\ }_{\mathrm{M}}$.
The task of enumerating
all linearly-independent high-gradient operators%
~(\ref{eq: def HGO if M,k to infty})
of order $s$ is greatly simplified by the assumption
$M,k\to\infty$.
We can rule out the scenario by which 
it is a finite set of independent Casimir operators
of gl($M|M$) that fixes all the linearly independent classical
high-gradient operators of order $s$ once the limit $M\to\infty$
has been taken.
We can also rule out the scenario by which normal
ordering changes the book-keeping between classical and
quantum high-gradient operators of order $s$ once the limit $k\to\infty$
has been taken.

For high-gradient operators of 
type Eq.\ (\ref{eq: classical hgo gl(M|N)k})
or (\ref{eq: def HGO if M,k to infty}), 
the coupling $g^{\ }_{\mathrm{A}}$ does not
renormalize their scaling dimensions, 
since $g^{\ }_{\mathrm{A}}$
(or $\mathcal{O}^{\ }_{\mathrm{A}}$)
can be removed from 
the action (\ref{eq: glMN perturbed by two current-current int.})
by chiral transformation.
All that therefore is needed
to compute their one-loop scaling dimensions are their
OPE with the quadratic Casimir operator 
$\mathcal{O}^{\ }_{\mathrm{M}}$.
We will write the following expressions for the general case of
$\mathrm{GL}(M|N)$, and will set $M=N$ 
(i.e., the case of interest)
only in Eqs.%
~(\ref{eq: result in large M,N for max and min lambda}),
(\ref{eq: result in large M, k for min x(s) max}),
and
(\ref{eq: result in large M, k for min x(s) min}).
The required OPEs follow from
(a) the intra-trace formula
\begin{subequations}
\begin{equation}
\begin{split}
&
\mathrm{str}\,
\big[
\Wick{24mm}J\mathcal{M} 
\Wickunder{20mm}\bar{J}\mathcal{N}
\big]
\times \mathrm{str}\,
\big[
J\bar{J}
\big]
\\
&
\quad
=
\mathrm{str}\,\left(J\mathcal{N}\right)
\mathrm{str}\,\left(\mathcal{M}\bar{J}\right)
-
\mathrm{str}\,\left(
\mathcal{M}
\right)
\mathrm{str}\,\left(
J\bar{J}\mathcal{N}
\right)
\\
&
\qquad
-
\mathrm{str}\,\left(
J\mathcal{M}\bar{J}
\right)
\mathrm{str}\,\left(
\mathcal{N}
\right)
+
\mathrm{str}\,\left(J\mathcal{M}\right)
\mathrm{str}\,\left(\bar{J}\mathcal{N}\right)
\end{split}
\label{eq: glMN large MN intra trace formula}
\end{equation}
and (b) the inter-trace formula
\begin{equation}
\begin{split}
&
\mathrm{str}\,
\big[
\Wick{32mm}J\mathcal{M} 
\big]
\mathrm{str}\,
\big[
\Wickunder{20mm}\bar{J}\mathcal{N}
\big]
\times \mathrm{str}\,
\big[
J\bar{J}
\big]
\\
&
\quad
=
\mathrm{str}\,\left(
J \mathcal{N}\bar{J}\mathcal{M}
\right)
-
\mathrm{str}\,
\left(
J\bar{J}
\mathcal{N}
\mathcal{M}
\right)
\\
&
\qquad\
-
\mathrm{str}\,\left(
J\mathcal{M}\mathcal{N}\bar{J}
\right)
+
\mathrm{str}\,
\left(
J
\mathcal{M}
\bar{J}
\mathcal{N}
\right)
\end{split}
\label{eq: glMN large MN inter trace formula}
\end{equation}
\end{subequations}
with $\mathcal{M}$ and $\mathcal{N}$ arbitrary operators.
Here we have used the short-hand notation of
Eq.\ (\ref{eq: def short hand OPE}).

To proceed we also need to distinguish linearly independent
high-gradient operators of order $s$. To this end,
a ``quantum number'', the number of switches, is introduced.%
~\cite{Lerner90}$^{-}$\cite{Mall93}
The number of switches of type 
$n^{\ }_{\uparrow}$ 
and of type
$n^{\ }_{\downarrow}$ 
in a single trace are defined as follows. Consider the trace
\begin{subequations}
\begin{equation}
\mathrm{str}\,
\big(
J^{\ }_{\mu^{\ }_{1}}
J^{\ }_{\mu^{\ }_{2}}
J^{\ }_{\mu^{\ }_{3}}
\cdots
J^{\ }_{\mu^{\ }_{2n}}
\big)
\label{eq: str needed to define number switches}
\end{equation}
where $\mu^{\ }_{1},\cdots,\mu^{\ }_{2n}=\pm$ 
while $J^{\ }_{-}=J$
and $J^{\ }_{+}=\bar{J}$.
Write the sequence of ``conformal'' indices
\begin{equation}
\mu^{\ }_{1},\cdots,\mu^{\ }_{2n},\mu^{\ }_{2n+1}
\end{equation}
\end{subequations}
where $\mu^{\ }_{2n+1}=\mu^{\ }_{1}$
by cyclicity of the trace. 
The number 
$n^{\ }_{\uparrow}$ 
of switches of type $\uparrow$ 
is the number of sign changes from 
$+\to -$ 
in two consecutive conformal indices when reading the sequence
$\mu^{\ }_{1},\cdots,\mu^{\ }_{2n},\mu^{\ }_{2n+1}$
from left to right.
The number 
$n^{\ }_{\downarrow}$ 
of switches of type $\downarrow$ 
is the number of sign changes from 
$-\to +$ 
in two consecutive conformal indices when reading the sequence
$\mu^{\ }_{1},\cdots,\mu^{\ }_{2n},\mu^{\ }_{2n+1}$
from left to right.

These quantum numbers are useful as it can be shown that
there is no contribution in the one-loop RG
of supertraces made out
of $2n$ currents as in
Eq.~(\ref{eq: str needed to define number switches})
from the subspace 
with $n^{\ }_{\uparrow}$ and $n^{\ }_{\downarrow}$ 
to the one with
at least $n^{\ }_{\uparrow}+1$ and $n^{\ }_{\downarrow}+1$.
This implies a lower triangular structure for the linearized RG equations
obeyed by all supertraces of order $2n$ as in
Eq.~(\ref{eq: str needed to define number switches})
which allows to treat separately each sector defined by
a given number of switches.
We shall assume that the strongest renormalization of the
engineering scaling dimensions occurs within the sector made
of the maximum number of switches.

Within the subspace of maximal switches it is sufficient to introduce
\begin{equation}
\omega:=
J \bar{J}\equiv
J^{\ }_{-}J^{\ }_{+},
\qquad
\Omega^{\ }_{m}:=
\mathrm{str}\,\big(\omega^{m}\big),
\end{equation}
for any $m=1,2,3,\cdots$.
\begin{subequations}
With the help of the OPE 
(\ref{eq: glMN large MN intra trace formula})
and
(\ref{eq: glMN large MN inter trace formula})
one verifies the OPE
\begin{equation}
\begin{split}
&
\mathrm{str}\,(\Wick{20mm}\omega \omega^m \Wickunder{13mm}\omega \omega^n) 
\times \mathcal{O}^{\ }_{\mathrm{M}}=
-\Omega^{\ }_{m+2}\Omega^{\ }_n
-\Omega^{\ }_{n+2}\Omega^{\ }_m
\\
&
\hphantom{
\mathrm{str}\,(\Wick{20mm}\omega \omega^m \Wickunder{13mm}\omega \omega^n) 
\times \mathcal{O}^{\ }_{\mathrm{M}}=
         }
-2\Omega^{\ }_{m+1}\Omega^{\ }_{n+1},
\\
&
\mathrm{str}\,(\Wick{28mm}\omega \omega^m)
\mathrm{str}\,(\Wickunder{13mm}\omega \omega^n)
\times 
\mathcal{O}^{\ }_{\mathrm{M}}=
-4\Omega^{\ }_{m+n+2},
\\
&
\mathrm{str}\,
(\Wick{13mm}\Wickunder{13mm}\omega \omega^m)
\times 
\mathcal{O}^{\ }_{\mathrm{M}}=
-\Omega^{\ }_1 \Omega^{\ }_m
-(N-M)\Omega^{\ }_{m+1},
\end{split}
\label{eq: OPE within maximal switch subspace}
\end{equation}
and
\begin{equation}
\begin{split}
&
\Wick{10mm}\Wickunder{10mm}\Omega^{\ }_m 
\times \mathcal{O}_{\mathrm{M}}
=
-2m \sum_{k,l=1}^{k+l=m}\Omega^{\ }_{k}\Omega^{\ }_{l}
-2m (N-M) \Omega^{\ }_m,
\\
&
\Wick{17mm}\Omega^{r_m}_m \Wickunder{11mm}
\Omega^{r_n}_n
\times 
\mathcal{O}_{\mathrm{M}}
=
-4 r_m r_n mn \Omega^{\ }_{m+n}
\Omega^{r_m-1}_m \Omega^{r_n-1}_n,
\label{eq: OPE within maximal switch subspace 2}
\end{split}
\end{equation}
for any 
$m,n,r^{\ }_m,r^{\ }_n=1,2,3,\cdots$.
\label{eq: OPE within maximal switch subspace all}
\end{subequations}

The action of the linearized one-loop RG flow
on the space of composite operators 
in the subspace of maximal switches
spanned by
\begin{eqnarray}
\Omega^{r^{\ }_{1}}_{1}
\Omega^{r^{\ }_{2}}_{2}
\cdots
\Omega^{r^{\ }_{L}}_{L},
\quad
\sum_{p=1}^{L} p \  r^{\ }_p=2s,
\end{eqnarray}
is encoded by the operator
\begin{eqnarray}
\hat{R}
\!\!&:=&\!\!\!
-2
\left(N-M\right)
\sum_{k}
k \Omega^{\ }_{k}\frac{\partial}{\partial \Omega^{\ }_{k}}
\nonumber\\
&&\!\!\!{}
-
2
\sum_{l,n}
\left[
(l+n)
\Omega^{\ }_{l} \Omega^{\ }_{n}
\frac{\partial}{\partial \Omega^{\ }_{l+n}}
+
ln\,
\Omega^{\ }_{l+n}
\frac{\partial}{\partial \Omega^{\ }_{l}}
\frac{\partial}{\partial \Omega^{\ }_{n}}
\right].
\nonumber\\&&
\label{eq: RG equation}
\end{eqnarray}
It is instructive to compare the OPE
(\ref{eq: OPE within maximal switch subspace all})
and the RG equation
(\ref{eq: RG equation})
with the corresponding result in the weakly coupled 
NL$\sigma$M on 
the symmetric space
$\mathrm{U}(P+Q)/\mathrm{U}(P)\times \mathrm{U}(Q)$
with $P,Q>1$:%
~\cite{Lerner90,Wegner91}
They 
are essentially identical to the corresponding result for the 
$\mathrm{U}(P+Q)/\mathrm{U}(P)\times \mathrm{U}(Q)$
NL$\sigma$M.

Now we return to the case $M=N$.
The diagonalization of $\hat{R}$ 
gives the largest and smallest eigenvalues%
~\cite{Lerner90, Wegner91}
\begin{eqnarray}
\lambda^{(s)}_{\mathrm{max}}=
+2 s(s-1)=
-\lambda^{(s)}_{\mathrm{min}}.
\label{eq: result in large M,N for max and min lambda}
\end{eqnarray}
Thus, both largest and smallest eigenvalues
depend quadratically on $s$.
In turn, one obtains a spectrum of one-loop scaling
dimensions with the upper and lower bounds
\begin{eqnarray}
&&
x^{(s)}_{\mathrm{max}}=
2s
+
\frac{g^{\ }_{\mathrm{M}}}{\pi}s(s-1),
\label{eq: result in large M, k for min x(s) max}
\\
&&
x^{(s)}_{\mathrm{min}}=
2s
-
\frac{g^{\ }_{\mathrm{M}}}{\pi}s(s-1),
\label{eq: result in large M, k for min x(s) min}
\end{eqnarray}
for any given $1<s=2,3,\cdots$. 
Observe that these bounds
are interchanged when 
$g^{\ }_{\mathrm{M}}\to-g^{\ }_{\mathrm{M}}$.

\subsection{
High-gradient operators
when $k=1$
           }
\label{subsec: largest HGO order s in GL(N|N) at k=1}

Having dealt with the extreme ``classical'' limit, 
we turn our attention to the extreme ``quantum'' limit
$M=1,2,3,\cdots$ and $k=1$ for which
the interacting field theory 
(\ref{eq: glMN perturbed by two current-current int.})
describes a problem of Anderson localization in $d=2$ dimensions
reviewed in Appendix~\ref{sec: HWK model}.

The classification of all independent
high-gradient operators in GL($M|M$)
or in PSL($M|M$)
is more involved than in SU(2)
because the problem of listing all invariants
is more complex.\cite{Bershadsky99}
An increase of complexity can already be 
seen at the level of SU($N$)
for which the invariant tensors of rank $2s$ are obtained
from all possible products of one rank 2 tensor and two rank 3 tensors.%
\cite{Dittner72}
Instead of considering the most generic family of ``classical''
high-gradient operators%
~(\ref{eq: classical hgo gl(M|N)k}),
we consider 
the GL($M|M$) invariant family of ``classical'' objects
\begin{equation}
\left\{
\left.
\mathcal{O}^{m}_{\mathrm{M}}
\mathcal{O}^{n}_{\mathrm{A}}
\right|
m,n=0,1,2,3,\cdots,
\quad
m+n=s
\right\},
\label{eq: hgo glMNk=1}
\end{equation}
which must then be normal ordered.
We are going to prove that the coupling constant $Z^{(s)}_{m,n}$
of the high-gradient operator
$\mathcal{O}^{m}_{\mathrm{M}}\mathcal{O}^{n}_{\mathrm{A}}$
in the action
\begin{equation}
\begin{split}
S=& 
S^{\ }_{*} 
+ 
\int \frac{d\bar zdz}{2\pi{i}} 
\left(
\frac{g^{\ }_{\mathrm{A}}}{2\pi} 
\mathcal{O}^{\ }_{\mathrm{A}}
+
\frac{g^{\ }_\mathrm{M}}{2\pi} 
\mathcal{O}^{\ }_{\mathrm{M}}
\right)
\\
&
-
\sum_{m,n = 0}^{m+n=s}
Z^{(s)}_{m,n}\mathfrak{a}^{2s-2}
\int \frac{d\bar zdz}{2\pi{i}} 
\mathcal{O}^{m}_{\mathrm{M}}
\mathcal{O}^{n}_{\mathrm{A}}
\end{split}
\end{equation}
obeys the linearized one-loop RG equation
\begin{equation}
\begin{split}
\frac{d Z^{(s)}_{m,n}}{dl}=&
\left(
2
-
2s
\right)
Z^{(s)}_{m,n}
\\
&
- 
4\frac{g^{\ }_{\mathrm{M}}}{2\pi}
m(m-1)Z^{(s)}_{m,n}
\\
&
+4 \frac{g^{\ }_{\mathrm{M}}}{2\pi}
(m+1)^2
Z^{(s)}_{m+1,n-1}
\label{eq: RG equations of Zmn in gl(MM)k=1}
\end{split}
\end{equation}
for any $m,n=0,1,2,3,\cdots$ with $m+n=s>1$.
For the $PSL(M|M)$  theory the operators $\mathcal{O}^{n}_{A}$
are all absent.

The RG equation (\ref{eq: RG equations of Zmn in gl(MM)k=1})
shows that there is no feedback from 
high-gradient operators containing a factor
$\mathcal{O}^{n}_{\mathrm{A}}$
to those containing a factor 
$\mathcal{O}^{n'}_{\mathrm{A}}$
provided $n'<n$.
Diagonalization of the RG equation gives
the set of one-loop scaling dimensions
\begin{equation}
\begin{split}
&
x^{(s)}_{m,n}=
2s
+
\frac{2g^{\ }_{\mathrm{M}}}{\pi}
m(m-1)
\end{split}
\label{eq: scaling dimension OM raised to the power s}
\end{equation}
for all $m,n=0,1,2,3,\cdots$ such that $m+n=s$.
For a positive 
$g^{\ }_{\mathrm{M}}$ we get the lower and upper bounds
\begin{equation}
x^{(s)}_{\mathrm{min}}=
2s,
\qquad
x^{(s)}_{\mathrm{max}}=
2s
+
\frac{2g^{\ }_{\mathrm{M}}}{\pi}
s(s-1),
\label{eq: scaling dimension OM raised to the power s if positive}
\end{equation}
respectively, i.e.,
$x^{(s)}_{m,n}$ with $m+n=s$
is always much larger than the engineering dimension
$2s$ so that the high-gradient operator 
$\mathcal{O}^{m}_{\mathrm{M}}\mathcal{O}^{n}_{\mathrm{A}}$
is irrelevant.
For a negative
$g^{\ }_{\mathrm{M}}$,
the spectrum of lower bounds on
$x^{(s)}_{m,n}$ with $m+n=s$
is unbounded from below when $s\to\infty$,
i.e., 
\begin{equation}
x^{(s)}_{\mathrm{min}}=
2s
-
\frac{2|g^{\ }_{\mathrm{M}}|}{\pi}
s(s-1),
\qquad
x^{(s)}_{\mathrm{max}}=
2s.
\label{eq: scaling dimension OM raised to the power s if negative}
\end{equation}

\textit{Proof:}
Having made the simplification
$g^{\ }_{\mathrm{A}}=0$
we only need to compute the OPE 
$
\mathcal{O}^{m}_{\mathrm{M}}
\mathcal{O}^{n}_{\mathrm{A}}
\times
\mathcal{O}_{\mathrm{M}}
$,
where $1\le m+n=s$, to justify
Eqs.%
~(\ref{eq: RG equations of Zmn in gl(MM)k=1})
and%
~(\ref{eq: scaling dimension OM raised to the power s}).
Each operator in Eq.%
~(\ref{eq: hgo glMNk=1})
contains terms with
$4s$ bosons,
$4s-2$  bosons and $2$ different fermions,
$4s-4$  bosons and $4$ different fermions,
...,
$4s-2M$ bosons and $2M$ different fermions,
and so on.
The terms that contain identical
fermions have short-distance singularities
and hence they should be interpreted as 
operators that involve gradients over fermion fields
after normal ordering.
It is understood from now on that the OPE
$
\mathcal{O}^{m}_{\mathrm{M}}
\mathcal{O}^{n}_{\mathrm{A}}
\times
\mathcal{O}_{\mathrm{M}}
$
is only over the terms in the expansion
$
\mathcal{O}^{m}_{\mathrm{M}}
\mathcal{O}^{n}_{\mathrm{A}}
$
involving different fermions, i.e., the OPE we present are
``accurate'' up to terms involving gradients over fermionic spinors.
Neglecting the OPE between derivatives
of the fermionic spinors and
$
\mathcal{O}^{\ }_{\mathrm{M}}
$
is harmless insofar as these OPE cannot feedback into the RG
flows of those contributions that we keep.

Let
\begin{equation}
(\chi \xi):=
\sum_{A=1}^{2M} 
\chi^A \xi^{\ }_A=
\sum_{A=1}^{2M} 
(-)^{A}
\xi^{\ }_A\chi^A 
\end{equation}
and remember that
$
\mathcal{O}_{\mathrm{A}} 
= -\left(\psi^{\dag}\psi\right)
\left(\bar{\psi}^{\dag}\bar{\psi}\right)
$
while
$
\mathcal{O}_{\mathrm{M}} 
= -\left(\psi^{\dag}\bar{\psi}\right)
\left(\bar{\psi}^{\dag}\psi\right)
$.}
The OPE that involve
$\big(\psi^{\dag}\bar{\psi}\big)$
and
$\big(\bar{\psi}^{\dag}\psi\big)$
are 
\begin{equation}
\begin{split}
&
\big(\Wick{18mm}\psi^{\dag}\Wickunder{10mm}\bar{\psi}\big) 
\times 
\big(\bar{\psi}^{\dag}\psi\big)=
0,
\\
&
(\Wick{26mm}\psi^{\dag}\bar{\psi})(\psi^{\dag}\Wickunder{9mm}\bar{\psi})
\times
(\bar{\psi}^{\dag}\psi)=
-
(\psi^{\dag} \bar{\psi}),
\\
&
(\Wick{35mm}\psi^{\dag}\bar{\psi})(\Wickunder{17mm}\bar{\psi}^{\dag} \psi)
\times
(\psi^{\dag}\bar{\psi})(\bar{\psi}^{\dag}\psi)=
-
\mathcal{O}^{\ }_{\mathrm{A}}.
\end{split}
\label{eq: OPE within the O_M sector}
\end{equation}
On the other hand,
the OPE that involve
$\big(\psi^{\dag} \psi \big)$,
$\big(\bar{\psi}^{\dag}\bar{\psi} \big)$,
$\big(\psi^{\dag}\bar{\psi}\big)$,
and
$\big(\bar{\psi}^{\dag}\psi \big)$ are given by
\begin{subequations}
\begin{eqnarray}
&&
  (\Wick{35mm}\psi^{\dag}\psi)
  (\Wickunder{17mm}\bar{\psi}^{\dag}\bar{\psi})
  \times
  (\psi^{\dag}\bar{\psi})(\bar{\psi}^{\dag}\psi)
\label{eq: OPE outside the O_M sector a}
\\
&&
\qquad
=
-
  (\Wick{35mm}\psi^{\dag}\psi)
  (\bar{\psi}^{\dag}\Wickunder{18mm}\bar{\psi})
  \times
  (\psi^{\dag}\bar{\psi})(\bar{\psi}^{\dag}\psi)
  =
-
\mathcal{O}^{\ }_{\mathrm{M}},
\nonumber\\
&&
  (\Wick{35mm}\psi^{\dag}\psi)
  (\Wickunder{17mm}\bar{\psi}^{\dag} \psi)
  \times
  (\psi^{\dag}\bar{\psi})(\bar{\psi}^{\dag}\psi)
\label{eq: OPE outside the O_M sector b}
\\
&& 
\qquad
=
  -(\psi^{\dag}\Wick{18mm}\psi)
  (\Wickunder{17mm}\bar{\psi}^{\dag} \psi)
  \times
  (\psi^{\dag}\bar{\psi})(\bar{\psi}^{\dag}\psi)
  =
  (\psi^{\dag}\psi)(\bar{\psi}^{\dag}\psi).
  \nonumber 
\end{eqnarray}
\label{eq: OPE outside the O_M sector}
\end{subequations}

Both $\mathcal{O}^{\ }_{\mathrm{M}}$ and $\mathcal{O}^{\ }_{\mathrm{A}}$
are generated
through the OPE
(\ref{eq: OPE within the O_M sector})
and
(\ref{eq: OPE outside the O_M sector}),
respectively.
However, 
two OPE
in 
Eq.\ (\ref{eq: OPE outside the O_M sector a})
always appear in a pairwise fashion
and cancel each other,
\begin{equation}
  (\Wick{25mm}\psi^{\dag}\psi)
  (\Wickunder{16mm}\bar{\psi}^{\dag}\bar{\psi})
  \mathcal{A}
  \times
  \mathcal{O}^{\ }_{\mathrm{M}}
  +
  (\Wick{25mm}\psi^{\dag}\psi)
  (\bar{\psi}^{\dag}\Wickunder{12mm}\bar{\psi})
  \mathcal{A}
  \times
  \mathcal{O}_{\mathrm{M}}=0,
\end{equation}
where $\mathcal{A}$ is some operator.
Hence, the total number of $\mathcal{O}^{\ }_\mathrm{M}$
contained in a high-gradient operator
never increases under the linearized RG flow.

{}From the OPE (\ref{eq: OPE within the O_M sector}) 
and (\ref{eq: OPE outside the O_M sector}) 
one deduces the OPE 
\begin{eqnarray}
&&
\mathcal{O}^{m}_{\mathrm{M}}
\mathcal{O}^{n}_{\mathrm{A}}
  \times
\mathcal{O}^{\ }_{\mathrm{M}}
\nonumber\\
&&\quad
=
m 
\mathcal{O}^{m-1}_{\mathrm{M}}
\mathcal{O}^{n}_{\mathrm{A}}
\Wick{10mm}
\Wickunder{10mm}
\mathcal{O}_{\mathrm{M}}
\times
\mathcal{O}_{\mathrm{M}}
+
n 
\mathcal{O}^{m}_{\mathrm{M}}
\mathcal{O}^{n-1}_{\mathrm{A}}
\Wick{10mm}
\Wickunder{10mm}
\mathcal{O}_{\mathrm{A}}
\times
\mathcal{O}_{\mathrm{M}}
\nonumber\\
&&\qquad
+
mn 
\mathcal{O}^{m-1}_{\mathrm{M}}
\mathcal{O}^{n-1}_{\mathrm{A}}
\Wick{15mm}
\mathcal{O}_{\mathrm{M}}
\Wickunder{10mm}
\mathcal{O}_{\mathrm{A}}
\times
\mathcal{O}_{\mathrm{M}}
\nonumber\\
&&\qquad
+
m(m-1) 
\mathcal{O}^{m-2}_{\mathrm{M}}
\mathcal{O}^{n}_{\mathrm{A}}
\Wick{16mm}
\mathcal{O}_{\mathrm{M}}
\Wickunder{10mm}
\mathcal{O}_{\mathrm{M}}
\times
\mathcal{O}_{\mathrm{M}}
\nonumber\\
&&\qquad
+
n(n-1) 
\mathcal{O}^{m}_{\mathrm{M}}
\mathcal{O}^{n-2}_{\mathrm{A}}
\Wick{15mm}
\mathcal{O}_{\mathrm{A}}
\Wickunder{10mm}
\mathcal{O}_{\mathrm{A}}
\times
\mathcal{O}_{\mathrm{M}}
\nonumber\\
&&\quad
=
2m(m-1) 
\mathcal{O}^{m}_{\mathrm{M}}
\mathcal{O}^{n}_{\mathrm{A}}
-
2m^2
\mathcal{O}^{m-1}_{\mathrm{M}}
\mathcal{O}^{n+1}_{\mathrm{A}}.
  \label{eq: full OPE hgo}
\end{eqnarray}
(When $m=0$, the term with 
$\mathcal{O}^{m-1}_{\mathrm{M}}$
is absent from the last line.)
The linearized one-loop RG equation%
~(\ref{eq: RG equations of Zmn in gl(MM)k=1})
thus follows
from the OPE (\ref{eq: full OPE hgo}).
$\square$

Had we assumed the level $k$ to be larger than $k=1$,
the family~(\ref{eq: hgo glMNk=1})
would not have been closed under the OPE with
$\mathcal{O}^{\ }_{\mathrm{M}}$.
For example, in the extreme
classical limit $M,k\to\infty$ 
the family of high-gradient operators is given by the much
larger family~(\ref{eq: def HGO if M,k to infty}).

We close by pointing out that we could have reached the same conclusions
on the spectrum of one-loop scaling dimensions of high-gradient operators
had we used, instead of the effective action with diagonal GL($M|M)$ 
symmetry, an action built out of fermionic replicas or an action built out
of bosonic replicas and taken the number of replicas to zero at the end of
the day. Using bosonic replicas mimics very closely the line
of argument presented here. Using fermionic replicas singles out 
high-gradient operators made of fermionic spinors that are all 
distinct through their replica index and then taking this replica index 
to zero, very much in the same way as replicated
vortices in certain classes of classical random
two-dimensional Coulomb gases.%
~\cite{Korshunov93}$^{-}$\cite{Fukui02}

We would like to stress that our 
results depend crucially on the 
continuous symmetry GL($2N|2N)$ 
of the $\widehat{\mathrm{gl}}(2N|2N)^{\ }_{k=1}$ Thirring model.
(From the point of view of Anderson localization,
it is the existence of a continuous symmetry
not the symmetry group per se that matters since 
the symmetry group changes depending on the choice made to
represent single-particle Green's functions, say
a supersymmetric, bosonic replicas, fermionic replicas, or Keldysh
path integral.)
If we consider local perturbations (local operators) 
that break the GL$(2N|2N)$ symmetry,
an infinite set of local operators with relevant (negative)
scaling dimensions can appear.
This alternative set of local operators may be related to
the situation recently considered by Le Doussal and Schehr.%
\cite{LeDoussal06}
The microscopic starting point of Ref.~\onlinecite{LeDoussal06}
is a class of classical random $XY$ models in two dimensions. 
These models can also be viewed as
interacting models of Dirac fermions subjected to disorder,
by the magic of the boson-fermion duality in 
$d=(1+1)$ dimensions.%
\cite{Mudry99,Guruswamy00} 
The difference with our paper is that 
their model is not invariant under a
\textit{continuous} symmetry group, but only under
the discretey symmetry group which permutes
the replica indices.
It is then necessary to use the full machinery of functional RG
to account for the one-loop relevance of high-gradient operators.

\subsection{Comparison with the 
$\mathbb{C}P^{1|2}$ NL$\sigma$M}

The perturbed
$\widehat{\mathrm{gl}}(2N|2N)^{\ }_{k=1}$ 
WZW model with $g^{\ }_{\mathrm{M}}>0$ (Thirring model)
describes a problem of Anderson localization in 
two dimensions.
As briefly reviewed in
Appendix~\ref{sec: HWK model}, 
this problem of Anderson localization 
arises as the long-wavelength description of
a tight-binding model on a two-dimensional
bipartite lattice with a form of disorder that
preserves sublattice and time-reversal symmetries.
The long-wavelength theory is a $(2+1)$-dimensional
Dirac equation subject to disorder potentials
consistent with these symmetries.
In terms of the symmetry-based classification of Anderson 
localization, the relevant symmetry class is the class BDI
(chiral-orthogonal symmetry class).%
~\cite{Zirnbauer96}$^{-}$\cite{Heinzner05}

It is possible to use a different 
representation of this Anderson localization problem, in terms of
a NL$\sigma$M
whose target space is the \textit{non-compact} supermanifold 
\begin{equation}
\mathrm{GL}(2N|2N)/\mathrm{OSp}(2N|2N).
\label{eq: NLSM for BDI}
\end{equation}
(A suitable analytical continuation in the boson-boson sector is needed 
to implement the non-compactness.~\cite{Zirnbauer96})
These two descriptions, one in terms of the 
Thirring model and the other in terms of the 
NL$\sigma$M, 
are complementary to each other in
that when  one of the models is strongly coupled,
the other is weakly coupled.
A reflection of this appears in the conductivity. The
coupling constant of the 
NL$\sigma$M is inversely proportional to the conductivity.
In the clean limit $g^{\ }_{\mathrm{M}}=0$ of the Thirring model
the conductivity is of order unity (in units of $e^2/h$),
consistent with the strongly coupled regime of the NL$\sigma$M.
The conductivity increases with $g^{\ }_{\mathrm{M}}>0$
as seen in perturbation theory.\cite{Ostrovsky06}
Furthermore,  both $g^{\ }_{\mathrm{M}}$ and the conductivity 
are exactly marginal. 
This suggests a deeper relationship between
the Thirring model and the NL$\sigma$M,
and indeed (following Ref.~\onlinecite{Ryu09}),
one can turn the Thirring model into the NL$\sigma$M continuously
by tuning $g^{\ }_{\mathrm{M}}$ (or equivalently the conductivity) continuously.  

We consider the case $N=1$ for which 
the \textit{non-compact} target supermanifold
(\ref{eq: NLSM for BDI}) 
is isomorphic to 
$\mathrm{U}(1)\times\mathrm{U}(1)\times\mathbb{C}P^{1|2}$,
where again a suitable analytical continuation is understood for
$\mathbb{C}P^{1|2}$, i.e.,
we need to consider the \textit{non-compact} counterpart to
$\mathbb{C}P^{1|2}$
as defined in Appendix~\ref{app: HGO on projective superspaces}.
Obtaining the non-compact $\mathbb{C}P^{1|2}$ target supermanifold
of the NL$\sigma$M from 
$\mathrm{U}(1)\times \mathrm{U}(1)\times\mathbb{C}P^{1|2}$
corresponds in the Thirring model to the reduction of
the $\mathrm{GL}(2|2)$ to the $\mathrm{PSL}(2|2)$ current algebra 
in Eq.\ (\ref{eq: PSL and GL}).

It is explicitly shown in Appendix%
~\ref{app: HGO on projective superspaces}
that all high-gradient operators are made more irrelevant 
at one-loop order by fluctuations
in any non-compact $\mathbb{C}P^{N+M-1|N}$ NL$\sigma$M
labeled by the non-negative integers $M$ and $N$.
To be more precise, 
we find that the largest and smallest
one-loop scaling dimensions
for the high-gradient operators of type%
~(\ref{eq: set of HGO}),
for a given $s$, are
\begin{equation}
\begin{split}
x^{(s)}_{\mathrm{max}}=&\,
2s 
+ 
2|t|s(s-1),
\\
x^{(s)}_{\mathrm{min}}=&\,
2s 
+ 
2|t|\times0, 
\end{split} 
\end{equation}
where $|t|>0$ is the coupling constant of the 
non-compact $\mathbb{C}P^{N-1|N}$ NL$\sigma$M.
This is fully consistent with our finding 
(\ref{eq: scaling dimension OM raised to the power s if positive})
in the Thirring model. 
We conclude that, in symmetry class BDI,  
high-gradient operators 
in the Thirring model with $g^{\ }_{\mathrm{M}}>0$
behave in the same way as 
in the corresponding NL$\sigma$M (i.e., the one that belongs
to the symmetry class BDI).

The sign of $g^{\ }_{\mathrm{M}}$
in the perturbed $\widehat{\mathrm{gl}}(2N|2N)^{\ }_{k=1}$ 
WZW model can be chosen to be negative,
$g^{\ }_{\mathrm{M}}<0$. 
If so, this field theory does not represent anymore
the moments of the single-particle Green's function
in a problem of Anderson localization in (bulk) two dimensions.
Nevertheless, this field theory does describe a problem of 
Anderson localization
which, however, now belongs to the \textit{different}
symmetry class CII (chiral-symplectic symmetry class)
describing the effect of disorder on the Dirac
fermions  which are known to appear
at the two-dimensional boundary of a three-dimensional 
topological band insulator in the same symmetry class.%
~\cite{Schnyder08,Ryu09,Hosur09} 
Equation~(\ref{eq: scaling dimension OM raised to the power s if negative})
implies that high-gradient operators are now made more relevant
by the current-current perturbation $g^{\ }_{\mathrm{M}}<0$
to one-loop order. 
As for the case with $g^{\ }_{\mathrm{M}}>0$,
a problem of Anderson localization in the symmetry class CII
is characterized by a NL$\sigma$M with a corresponding target
manifold. As before, the beta function of the coupling constant
$g^{\ }_{\mathrm{M}}<0$
of the Thirring model as well as that of the coupling
constant of the corresponding
NL$\sigma$M  vanish, and
one can interpolate\cite{Ryu09}
between the weak coupling limit of 
the Thirring model and the strong coupling limit
of the NL$\sigma$M
and conversely, by tuning $g^{\ }_{\mathrm{M}}$ continuously.
The target supermanifold in symmetry class CII is
the \textit{compact} supermanifold~(\ref{eq: NLSM for BDI}),
from which one extracts when $N=1$ the NL$\sigma$M
 on the
\textit{compact} supermanifold $\mathbb{C}P^{1|2}$.
\cite{CommentSchomerusSaleur2009}
It is explicitly shown in Appendix%
~\ref{app: HGO on projective superspaces}
that all high-gradient operators are made more relevant 
at one-loop order by fluctuations
in any compact $\mathbb{C}P^{N+M-1|N}$ NL$\sigma$M
labeled by the non-negative integers $M$ and $N$.
In particular, for $M=0$,
we find that the largest and smallest
one-loop scaling dimensions
for the high-gradient operators of type%
~(\ref{eq: set of HGO}),
for a given $s$, are
\begin{equation}
\begin{split}
x^{(s)}_{\mathrm{max}}=&\,
2s 
- 
2ts(s-1),
\\
x^{(s)}_{\mathrm{min}}=&\,
2s 
+ 
2t \times0,
\end{split} 
\end{equation}
where $t>0$ is the coupling constant of the 
compact $\mathbb{C}P^{N-1|N}$ NL$\sigma$M. 
Once again, we conclude that  
high-gradient operators behave in the same way
in the Thirring model with $g^{\ }_{\mathrm{M}}<0$
and in the corresponding NL$\sigma$M that belongs to the
symmetry class CII.

\section{Conclusions}
\label{sec: conclusion}

More than twenty years after their discovery, 
the role of  high-gradient operators, which appear
to be highly relevant in one-loop computations of anomalous
dimensions in a great variety of NL$\sigma$Ms,
still remains a puzzle.
Indeed, this  perturbative property is rather general
as it can apply to both compact and non-compact target manifolds.
In the absence of an exact calculation of observables that
would be sensitive to high-gradient operators, 
it is still an outstanding question 
whether the extreme
RG-relevance of these operators is an artifact of the
one-loop calculation (e.g., in the $2+\epsilon$-expansion), 
or is a feature that is generally valid.
(For an attempt to compare the $\epsilon$ 
expansion in $d=2-\epsilon$ dimensions
with exact results obtained for $d=1$, see Ref.~\onlinecite{Ryu07a}.)

In order to shed some light on these issues
we have asked in this paper the following question.
Can high-gradient operators become relevant in the family of
two-dimensional
$\widehat{\mathrm{gl}}(M|M)^{\ }_{k}$ Thirring models
with $M$ and $k$ positive integers?
The strategy that we followed has three steps.
The first step consists of identifying all
the independent ``classical'' high-gradient operators of order $s$.
This is a problem of group theory that involves the enumeration of
all distinct GL($M|M$) singlets 
in the direct product 
of $2s$ adjoint representations
of GL($M|M$). 
The second step consists of normal-ordering all independent
classical high-gradient operators of order $s$. This step depends
crucially on the level $k$ of the non-Abelian Thirring model.
The inverse level $1/k$ plays the role of a quantum parameter that
vanishes in the limit $k\to\infty$. The level $k=1$ is thus the most 
``quantum''.
The computation of the linearized RG flows for the high-gradient 
operators is the final step.

We could not solve the first step in its full generality. 
We were nevertheless able to construct two sets 
of high-gradient operators in the
extreme ``classical'' limit 
$\widehat{\mathrm{gl}}(M|M)^{\ }_{k}$ with $M,k\to\infty$
and the extreme ``quantum'' case 
$\widehat{\mathrm{gl}}(M|M)^{\ }_{k}$ with $M$ a positive integer and $k=1$,
respectively,
and carry out the second and third steps consistently,
i.e., show that each family
of normal-ordered high-gradient operators is closed under 
the linearized RG flow equations.

The set of high-gradient operators that we considered in the extreme
``quantum'' limit is much smaller than the set of
high-gradient operators for the extreme ``classical'' case.
This is to be expected as normal ordering is extremely sensitive
to the free-field fermionic representation of the 
$\widehat{\mathrm{gl}}(M|M)^{\ }_{k}$ 
current algebra at the unperturbed} WZW critical point.
This difference has dramatic consequences for the
spectrum of one-loop anomalous scaling dimensions
in the extreme ``classical'' and ``quantum'' cases.%
~\cite{Gepner86}

In the extreme ``classical'' case, 
anomalous  one-loop scaling dimensions 
for high-gradient operators of order $s$ are
distributed in a symmetric fashion about zero with 
the minimum and the maximum both depending
quadratically on the order $s$,
very much like for the family 
of NL$\sigma$Ms on the target spaces
$\mathrm{U}(M+N)/\mathrm{U}(M)\times\mathrm{U}(N)$
with $M$ and $N$ positive integers.%
~\cite{Ryu07a,Altshuler86a}$^{-}$\cite{Altshuler91}
Hence, high-gradient operators
must become (one-loop) relevant for both signs of the current-current 
interaction with increasing order $s$
very much in the same way as their cousins do in both
the compact family 
$\mathrm{U}(M+N)/\mathrm{U}(M)\times\mathrm{U}(N)$
and the non-compact family
$\mathrm{U}(M,N)/\mathrm{U}(M)\times\mathrm{U}(N)$
with $M,N>1$.

In the extreme quantum case $k=1$, the spectrum of
anomalous one-loop scaling dimensions
of order $s$ is always one-sided,
i.e., positive for one sign of the current-current interaction.
For $\widehat{\mathrm{gl}}(M|M)^{\ }_{k=1}$ with $M$ a positive integer
the sign of the current-current interaction for which high-gradient
operators are always irrelevant corresponds to the interpretation
of the $\widehat{\mathrm{gl}}(2N|2N)^{\ }_{k=1}$ Thirring model as 
a problem of Anderson localization
in random tight-binding models on two-dimensional
bipartite lattices (symmetry class BDI). We have shown in this paper that
the high-gradient operators in these random tight-binding models
are irrelevant at one-loop order.

High-gradient operators in those NL$\sigma$Ms 
of relevance to the physics of Anderson localization
are related to the moments of the $dc$ conductance.%
~\cite{Altshuler86a}$^{-}$\cite{Altshuler91}
Their perturbative one-loop relevance has been interpreted
as the signature of broad tails in the probability distribution
of the conductance in Refs.~%
\onlinecite{Altshuler86a}--\onlinecite{Altshuler91}. 
(One should, however, bear in mind that 
the current-current correlation function entering 
the Kubo formula for the conductance 
looks rather different from a simple 
$\mathrm{GL}(2N|2N)$ current-current 
correlation function.%
~\cite{Ryu07b})
It would thus be very interesting to study the
probability distribution of
the $dc$ conductance in the relevant random tight-binding model
using nonperturbative techniques 
(this may include, e.g., also numerical approaches)
in order to establish if it is broad or not. 

\section*{Acknowledgments}

CM would like to thank Eduardo Fradkin for important comments.
This research was supported in part by the National Science Foundation
under Grant No.\ PHY05-51164 and under Grant No.\ DMR-0706140 (AWWL).
SR thanks the Center for Condensed Matter Theory at University of California,
Berkeley for its support. 

\appendix

\section{ 
High-gradient operators in NL$\sigma$Ms
on the complex projective superspaces $\mathbb{C}P^{N+M-1|N}$
        }
\label{app: HGO on projective superspaces}

Whether or not the spectra
of anomalous one-loop scaling dimensions 
of high-gradient operators in NL$\sigma$Ms
are symmetric about zero or not
can be very important
when the analytical continuation of the  
coupling constant $t$ 
in the NL$\sigma$M from positive to negative values
is meaningful from a physical point of view.
We shall call spectra which are fully symmetric
about zero {\it two-sided} spectra.
Spectra of anomalous one-loop scaling dimensions
that are strictly positive (negative) will be called
one-sided. NL$\sigma$Ms with the target manifolds
$S^{N-1}=\mathrm{O}(N)/\mathrm{O}(N-1)$
are already known to be one-sided.~\cite{Wegner90}
We are going to show that this is also 
the case for the NL$\sigma$Ms with the target manifolds\cite{SuperCPN}
\begin{equation}
\begin{split}
&
\mathbb{C}P^{N+M-1|N}
\simeq
\\
&
\mathrm{U}(N+M|N)/
[\mathrm{U}(1)\times\mathrm{U}(N+M-1|N)].
\end{split}
\label{eq: def CP{N+M-1|N}}
\end{equation}
The complex projective superspaces%
~(\ref{eq: def CP{N+M-1|N}})
are generalizations of the compact complex projective spaces
\begin{equation}
\mathbb{C}P^{M-1}\simeq
\mathrm{U}(M)/
[\mathrm{U}(1)\times \mathrm{U}(M-1)].
\label{eq: def compact CP{M-1}}
\end{equation}
We shall also study on their own right
the high-gradient operators in NL$\sigma$Ms with
the non-compact complex projective target spaces
\begin{eqnarray}
\mathrm{U}(1, N+M-1|N)/[
\mathrm{U}(1)\times \mathrm{U}(N+M-1|N)].
\label{eq: def non-compact CP{M-1}}
\end{eqnarray}
These non-compact manifolds follow from 
the compact complex projective spaces%
~(\ref{eq: def compact CP{M-1}})
upon analytical continuation 
of some real coordinates to imaginary ones.
The complex projective superspaces%
~(\ref{eq: def CP{N+M-1|N}})
are special cases of the K\" ahler supermanifolds 
whose high-gradient operators were studied
in Ref.~\onlinecite{Ryu07a}. We refer the reader for
notations, conventions, and the relevant intermediary results to
Ref.\ \onlinecite{Ryu07a}.

Appendix~\ref{app: HGO on projective superspaces} 
is organized as follows.
We first define NL$\sigma$Ms
on the projective supespaces~(\ref{eq: def CP{N+M-1|N}})
using a geometrical approach. We then present the 
main result of Appendix~\ref{app: HGO on projective superspaces}
on the one-loop scaling dimensions of high-gradient operators
and show that they are one sided. We close 
by briefly outlining how the
one-loop scaling dimensions of high-gradient operators
are computed.

\subsection{
Geometry of the $\mathbb{C}P^{N+M-1|N}$ NL$\sigma$Ms
           }
\label{appsubsec: geometry}

A NL$\sigma$M on a Hermitian supermanifold 
can be defined with the help of the partition function 
\begin{equation}
\begin{split}
Z:=&\,
\int \mathcal{D}
\left[\varphi^{\dag},\varphi\right]
e^{-S[\varphi^{\dag},\varphi]},
\\
S:=&\,
\frac{1}{2\pi t}
\int d^d r\,
\left(\partial^{\ }_{\mu} \varphi^{* a}\right)\,
{ }_{a^*}g_{b}(\varphi^{\dag},\varphi)\, 
\left(\partial^{\ }_{\mu} { }^b\varphi\right).
\end{split}
\label{eq: def hermitian nlsm}
\end{equation}
Here,
$(\varphi^{\dag}, \varphi)$ 
are the coordinates on the Hermitian supermanifold,
${ }_{a^*}g_{b}(\varphi^{\dag},\varphi)$
is the metric on the Hermitian supermanifold,
and $t$ is the NL$\sigma$M's coupling constant.

We are going to restrict our analysis to
Hermitian supermanifolds such that their metric 
can be derived from a K\" ahler potential. 
Furthermore, we shall choose the K\" ahler 
potential so that the target supermanifold
is none but the $\mathbb{C}P^{N+M-1|N}$
symmetric space.
The K\"ahler potential for $\mathbb{C}P^{N+M-1|N}$ is
\begin{subequations}
\label{eq: restriction Hermitean to CP manifold}
\begin{equation}
K:=\log(1+\varphi^{\dag}\xi\varphi).
\label{eq: def Kaehler for superCPn}
\end{equation}
The bilinear form
\begin{equation}
\varphi^{\dag}\xi\varphi:=
\varphi^{a*}\,
{}^{\vphantom{a}}_{a^*}\xi^{\ }_{b}\,
\varphi^{b}
\end{equation}
is presented in terms of the diagonal tensor $\xi$ 
with the components
\begin{equation}
{}^{\vphantom{a}}_{a^*}\xi^{\ }_{b}:= 
{}^{\vphantom{a}}_{a^*}\delta^{\ }_{b}
\end{equation} 
that do not depend on the coordinates
$(\varphi^{\dag},\varphi)$.
Hence,
\begin{equation}
1+\varphi^{\dag}\xi\varphi = 
1+
\sum_{i=1}^{N+M-1}
\phi^{*i} \phi^{i}
+
\sum_{i=1}^{N}
\psi^{*i}\psi^{i},
\end{equation}
\end{subequations}
where
$(\phi^{*i}, \phi^i)$ 
with ${i=1,\cdots,N+M-1}$ and
$(\psi^{*i}, \psi^i)$ with
${i=1,\cdots,N}$ 
are the bosonic and fermionic coordinates of
$(\varphi^{\dag}, \varphi)$,
respectively. We observe that
$(\varphi^{*a}, \varphi^a)$ has
$N+M-1$ bosonic and $N$ fermionic coordinates.
Equations~(\ref{eq: def hermitian nlsm})
and (\ref{eq: restriction Hermitean to CP manifold})
define the $\mathbb{C}P^{N+M-1|N}$ NL$\sigma$Ms.
Setting $N=0$ in
Eqs.~(\ref{eq: def hermitian nlsm})
and (\ref{eq: restriction Hermitean to CP manifold})
defines the NL$\sigma$Ms 
on the 
compact complex projective manifold%
~(\ref{eq: def compact CP{M-1}}).
The analytical continuation
$\phi^i\to i \phi^i$ 
and
$\phi^{*i}\to i \phi^{*i}$
in
Eqs.~(\ref{eq: def hermitian nlsm})
and (\ref{eq: restriction Hermitean to CP manifold})
defines the NL$\sigma$Ms on the non-compact 
complex projective manifold~(\ref{eq: def non-compact CP{M-1}}).

The derivative of the K\"ahler potential%
~(\ref{eq: def Kaehler for superCPn})
gives the metric
(a superanalogue of the Fubini-Study metric) 
for $\mathbb{C}P^{N+M-1|N}$ through
\begin{subequations}
\label{eq: metric for super CPn}
\begin{equation}
{}^{\vphantom{a}}_{a^*}g^{\vphantom{a}}_{b}=
\frac{\overrightarrow{\partial}}{\partial \varphi^{a*} }
K 
\frac{\overleftarrow{\partial}}{\partial \varphi^{b }}
\equiv
\mathsf{Z}\
{}^{\ }_{a^*}\mathsf{Y}^{\ }_{b}. 
\label{eq: Z and Y}
\end{equation}
We have introduced the scalar
\begin{equation}
\mathsf{Z}:=
\frac{
1
     } 
     {
1+\varphi^{\dag}\xi\varphi
     }
\label{eq: Z}
\end{equation}
and the tensor
\begin{equation}
{}^{\ }_{a^*}\mathsf{Y}^{\ }_{b}:=
{ }^{\ }_{a^*}\xi^{\ }_{b}
-
{ }^{\ }_{a^*}\xi^{\ }_c
\varphi^{c}
\mathsf{Z}
\varphi^{d*}
{ }^{\ }_{d^*}\xi^{\ }_b.
\label{eq: Y}
\end{equation}
\end{subequations}
Following the usage for graded indices
from Ref.~\onlinecite{Ryu07a},
we also define the tensors
\begin{equation}
\begin{split}
&
{}^{a}\mathsf{Y}^{b^*}:=
{}^{a}\xi^{b^*}
+
\varphi^{a}\varphi^{b*}, 
\\
&
{}^{\ }_{a}\mathsf{Y}^{\ }_{b^*}:=
(-1)^{a+b+ab}\,
{}_{b^* }^{\ }\mathsf{Y}_{a}^{\ },
\\
&
{}^{a^*}\mathsf{Y}^{b}:=
(-1)^{ab}\,
{}^{b}\mathsf{Y}^{a^*}.
\end{split}
\end{equation}
It then follows that
the metric indices can be raised, lowered, or shifted
according to
\begin{equation}
\begin{split}
&
{}^{\ }_{a^*}g^{\ }_{b}=
\mathsf{Z}\
 {}^{\ }_{a^*}Y^{\ }_{b},
\\
&
{}^{\ }_{a}g^{\ }_{b^*}=
\mathsf{Z}\ 
{}^{\ }_{a}\mathsf{Y}^{\ }_{b^*},
\\
&
{}^{a^*}g^{b}=
\mathsf{Z}^{-1} 
{}^{a^*}\mathsf{Y}^{b},
\\
&
{}^{a}g^{b^*}=
\mathsf{Z}^{-1}
{}^{a}\mathsf{Y}^{b^*}.
\end{split}
\end{equation}

The metric tensor~(\ref{eq: metric for super CPn}) 
can be expanded about any 
point of the manifold at which it is finite
(flat geometry). The lowest order in this
expansion defines the ``kinetic'' contribution
to the Lagrangian of the NL$\sigma$M,
whereas the higher-order contributions define the
``interactions''.
The bosonic contribution to the kinetic energy
must be positive definite for the path integral%
~(\ref{eq: def hermitian nlsm})
to be well defined. This condition fixes the sign
of the coupling constant $t$. 
For the compact complex projective manifolds%
~(\ref{eq: def compact CP{M-1}}),
$t>0$ must be chosen.
For the non-compact complex projective manifolds%
~(\ref{eq: def non-compact CP{M-1}}),
$t<0$ must be chosen
(see, for example, Ref.\ \onlinecite{Hof86}).
A consequence of the analytical continuation $t\to-t$
for the one-loop beta function of $t$,
if it is proportional to $t^2$ as it is in
$d=2$ dimensions, is that it changes 
by a sign. Similarly, the one-loop corrections 
to the scaling dimensions of high-gradient
operators also change by a sign under 
the analytical continuation $t\to-t$.%
~\cite{Polyakov01,Tseytlin03}
It then matters greatly whether the 
anomalous one-loop scaling dimensions
are two-sided or one-sided.

According to Friedan,%
~\cite{Friedan85}
the one-loop beta function for 
the coupling constant $t$ of a NL$\sigma$M on
a Riemannian manifold is given by the curvature
of the manifold. The curvature follows from the Ricci tensor,
which we now compute for 
$\mathbb{C}P^{N+M-1|N}$.
Needed are the coefficients of the connection.
They are
\begin{subequations}
\begin{equation}
\begin{split}
\Gamma^{a}_{\ b c}:=&\,
{}^{a}g^{d^*}\,
{}^{\ }_{d^*}g^{\ }_{b,c}
\\
=&\,
-\mathsf{Z}
\Big[
{}^{a}\delta^{\ }_{b}\,
\varphi^{d*}{}^{\ }_{d^*}\xi^{\ }_{c}
+
(-1)^{b c}\,
{}^{a}\delta^{\ }_{c}\,
\varphi^{d*}
{}^{\ }_{d^*}\xi^{\ }_{b}
\Big] 
\end{split}
\end{equation}
and
\begin{equation}
\begin{split}
\Gamma^{a^*}_{\ b^* c^*}=&\,
{}^{a^*}g^{d}\,
{}^{\ }_{d}g^{\ }_{b^*,c^*}
\\
=&\,
-\mathsf{Z}\,
\Big[
(-1)^{c}\,
{}^{a}\delta^{\ }_{b}\,
\varphi^{d}
{}^{\ }_{d}\xi^{\ }_{c^*}
+
(-1)^{b+bc}\,
{}^{a}\delta^{\ }_{c}\,
\varphi^{d}
{}^{\ }_{d}\xi^{\ }_{b^*}
\Big].
\end{split}
\end{equation}
\end{subequations}
The curvature tensor field on
$\mathbb{C}P^{N+M-1|N}$
can then be expressed solely in terms of
the metric tensor field,
\begin{subequations}
\label{eq: curvature yensor field super cpn}
\begin{equation}
\begin{split}
R^{a}_{\ b c d^*}=&\,
-
\Gamma^{a}_{\ bc}
\overleftarrow{\partial}^{*}_{d}
\\
=&\,
{}^{a}\delta^{\ }_{b}\,
g^{\ }_{c d^*}
+
{}^{a}\delta^{\ }_{c}\,
(-1)^{bc}\,
g^{\ }_{b d^* }
\end{split}
\end{equation}
and
\begin{equation}
\begin{split}
R^{a^*}_{\ b^* c^* d}=&\,
-\Gamma^{a^*}_{\ b^* c^* }
\overleftarrow{\partial}_{d}^{\ }
\\
=&\,
{}^{a}\delta^{\ }_{b}\,
g_{c^* d}^{\ }
+
{}^{a}\delta^{\ }_{c}\,
(-1)^{b c}\,
g^{\ }_{b^* d}.
\end{split}
\end{equation}
\end{subequations}
For $\mathbb{C}P^{N+M-1|N}$,
the Ricci tensor field is proportional to the metric
with $M$ the proportionality constant,
\begin{equation}
R^{\ }_{b d^*}=
M g^{\ }_{bd^*}.
\label{eq: Ricci}
\end{equation}
For $\mathbb{C}P^{N+M-1|N}$,
it follows that the Ricci tensors vanishes when $M=0$,
and so does the one-loop beta function according to Friedan.
The beta function vanishes to all orders in the loop expansion.%
~\cite{Gade91-93} The special case of $\mathbb{C}P^{1|2}$
[$(M,N)=(0,2)$]
has also been discussed in
Refs.~\onlinecite{SchomerusSaleurEtAl2009} and \onlinecite{Read01}.

\subsection{
High-gradient operators for the
$\mathbb{C}P^{N+M-1|N}$ NL$\sigma$Ms
           }
\label{subsec: High-gradient operators for the CP...}

{}From the property%
~(\ref{eq: curvature yensor field super cpn}),
i.e., that the curvature tensor field 
of the supermanifold $\mathbb{C}P^{N+M-1|N}$
depends solely on its metric,
follows that the RG equations 
among the infinite set of operators 
made of local polynomials in
\begin{equation}
\mathcal{G}^{\ }_{\mu\nu}:=
\partial^{\ }_{\mu}\varphi^{a*} 
{}^{\ }_{a^*}g^{\ } _{b}
\partial^{\ }_{\nu}{}^{b}\varphi
\label{eq: def building blocks}
\end{equation}
are closed.

Near two dimensions ($d=2+\epsilon$), it is convenient to use
the conformal coordinates,
\begin{equation}
\partial^{\ }_{\pm}= 
\partial^{\ }_x 
\pm{i} 
\partial^{\ }_y,
\qquad
\mu=\pm,\nu=\pm,
\end{equation}
i.e., we use 
$\mathcal{G}_{++}$,
$\mathcal{G}_{+-}$,
$\mathcal{G}_{-+}$,
and
$\mathcal{G}_{--}$
as the building blocks 
for the high-gradient operators.
It can be shown that the one-loop RG equations 
are closed within the family
\begin{eqnarray}
 \big\{
\mathcal{G}^{p}_{+-}
\mathcal{G}^{q}_{-+}
\left(
\mathcal{G}^{\ }_{++}
\mathcal{G}^{\ }_{--}
\right)^{r}
\big\}^{p+q+2r=s}_{p=0,q=0,r=0}
\label{eq: set of HGO}
\end{eqnarray}
of high-gradient operators for any 
given number of gradients $2s$,
where $p$, $q$, and $r$ are any non-negative integer satisfying 
$p+q+2r=s$.
Furthermore, for any given $s$, $r$, and $r'$
the family~(\ref{eq: set of HGO})
obeys one-loop RG equations with an upper triangular structure
in the sense that all high-gradient operators
of the form~(\ref{eq: set of HGO})
with $r'>r$ do not enter the one-loop RG equations
for those high-gradient operators with $r$ fixed.
The task of diagonalizing the closed
one-loop RG equations obeyed by 
the family~(\ref{eq: set of HGO}) 
thus simplifies greatly. It is indeed sufficient to fix
$s$ and $r$ and to diagonalize the 
one-loop RG equations obeyed by 
the family~(\ref{eq: set of HGO}) 
labeled by the non-negative integers $p$ and $q$.
For any finite order $s$,
diagonalization of the one-loop RG flows
obeyed by the family~(\ref{eq: set of HGO})
of high-gradient operators yields
the one-loop RG eigenvalues
\begin{subequations}
\label{eq: main result}
\begin{equation}
\alpha^{(s)}_{p,q,r}=
-2 M r
+2
\Big(
-pq
+p(p-1)
+q(q-1)
\Big),
\end{equation}
here labeled by the non-negative integers $q$, $q$, and $r$
that satisfy
\begin{equation}
p+q+2r=s.
\end{equation}
\end{subequations}
Combining Eq.~(\ref{eq: main result})
with the engineering scaling dimension $2s$
yields the one-loop scaling dimensions
\begin{equation}
\begin{split}
x^{(s)}_{p,q,r}=&\,
2s 
- 
t 
\alpha^{(s)}_{p,q,r}
\\
=&\,
2 s 
- 
2t 
\Big(
-
Mr
-
pq
+
p(p-1)
+
q(q-1)
\Big)
\end{split}
\label{eq: main result2}
\end{equation}
for the family~(\ref{eq: set of HGO})
of high-gradient operators.
Equations 
(\ref{eq: main result}) 
and 
(\ref{eq: main result2})
are the main result of this Appendix. 
Observe that this result is independent of 
the integer $N$ in $\mathbb{C}P^{N+M-1|N}$.
Hence, it applies to the case $N=0$, 
both in its compact and non-compact incarnations%
~(\ref{eq: def compact CP{M-1}})
and
(\ref{eq: def non-compact CP{M-1}}),
respectively.

We now take a closer look at the spectrum when $M=0$.
In this case, the projective superspace is Ricci flat, 
i.e., $R^{\ }_{a^* b}=0$
according to Eq.\ (\ref{eq: Ricci}),
and hence the one-loop beta function of 
the NL$\sigma$M coupling constant $t$ vanishes.
(These are \cite{Gade91-93} 
in fact lines of critical points
labeled by the coupling constant $t$ of the 
$\mathbb{C}P^{N-1|N}$ NL$\sigma$Ms.)
We also distinguish the compact case from the 
non-compact case by demanding that $t>0$
in the former case and that $t<0$ in the latter case.

The compact case corresponds to $t>0$.
For any given order $s$,
we seek the largest and smallest one-loop RG eigenvalues
that govern the RG flow of the high-gradient operators%
~(\ref{eq: set of HGO}) 
in the NL$\sigma$Ms 
$\mathbb{C}P^{N-1|N}$. 
Needed are the extremal values of
$\alpha^{(s)}_{p,q,r}$ 
while holding $p+q+2r=s$ fixed.
We find that the most and least dominant 
one-loop scaling dimensions in two dimensions
and for a fixed $s$ are
\begin{equation}
\begin{split}
x^{(s)}_{\mathrm{min}}=&\,
2s 
- 
2ts(s-1),
\\
x^{(s)}_{\mathrm{max}}=&\,
2s 
- 
2t\times0.
\end{split} 
\label{case M=0 compact}
\end{equation}
We conclude that the spectrum of one-loop anomalous scaling dimensions
(\ref{eq: main result}) 
for any ``compact''
$\mathbb{C}P^{N-1|N}$ NL$\sigma$M
is one-sided in the sense that 
it is not symmetrically distributed about zero:
While $x^{(s)}_{\mathrm{min}}$ is not bounded as a function of $s$,
$x^{(s)}_{\mathrm{max}}=2s$ irrespective of $s$. 
The result for the most dominant scaling dimension 
$x^{(s)}_{\mathrm{min}}$
is the same as that for the 
\begin{equation}
\mathrm{U}(P+Q)/\mathrm{U}(P)\times\mathrm{U}(Q)
\label{eq: U P+Q}
\end{equation}
NL$\sigma$Ms with $P,Q>1$.\cite{Lerner90,Ryu07a}
However,
the spectrum of one-loop anomalous scaling dimensions
for the NL$\sigma$Ms~(\ref{eq: U P+Q})
with $P,Q>1$ is two-sided:
The one-loop anomalous scaling dimensions 
are symmetrically distributed about zero.

The non-compact case corresponds to $t<0$.
For any given order $s$,
we seek the largest and smallest one-loop RG eigenvalues
that govern the RG flow of the high-gradient operators%
~(\ref{eq: set of HGO}) 
when $M=0$.
These follow from Eq.~(\ref{case M=0 compact})
with the substitution $t\to -t$,
\begin{equation}
\begin{split}
x^{(s)}_{\mathrm{max}}=&\,
2s 
+ 
2|t|s(s-1),
\\
x^{(s)}_{\mathrm{min}}=&\,
2s 
+ 
2|t|\times0. 
\end{split}
\label{case M=0 noncompact}
\end{equation}
So, there is no relevant high-gradient operator
in this non-compact case.
This is the consequence of the one-sided property of the spectrum
(\ref{eq: main result})
when $M=0$. On the other hand,
in the case of the non-compact 
\begin{equation}
\mathrm{U}(P,Q)/\mathrm{U}(P)\times \mathrm{U}(Q)
\end{equation}
NL$\sigma$Ms with $P,Q>1$, 
there are always relevant high-gradient operators.
We note that the one-loop scaling dimensions
(\ref{case M=0 compact}) and (\ref{case M=0 noncompact}) 
turn into the corresponding scaling dimensions
(\ref{eq: scaling dimension OM raised to the power s if negative})
and
(\ref{eq: scaling dimension OM raised to the power s if positive})
for the $\widehat{\mathrm{gl}}(2N|2N)^{\ }_{k=1}$ WZW model,
if we identify $t$ with $-g^{\ }_{\mathrm{M}}/\pi$.

We now relax the condition for criticality $M=0$
of the $\mathbb{C}P^{N-1|N}$
NL$\sigma$M target manifold.
It can then also be shown that the spectra~(\ref{eq: main result})
labeled by $s$ and $M$ are one-sided. 
Since 
this result is, as required, independent of $N$, it
applies to the $\mathbb{C}P^{M-1}$
NL$\sigma$Ms as well. In turn, 
$\mathbb{C}P^{M-1}
\sim 
\mathrm{U}(M)/\mathrm{U}(1)\times \mathrm{U}(M-1)
$ 
is obtained from 
$\mathrm{U}(P+Q)/\mathrm{U}(P)\times \mathrm{U}(Q)$
by specializing to $(P,Q)=(M-1,1)$ or $(1,M-1)$.
The reason why the spectrum of one-loop anomalous scaling dimensions in
$\mathrm{U}(P+Q)/\mathrm{U}(P)\times \mathrm{U}(Q)$
with $P,Q>1$ looks so different from the cases 
with either $P$ or $Q$ being unity is the following.
The $\mathrm{U}(P+Q)/\mathrm{U}(P)\times\mathrm{U}(Q)$
NL$\sigma$Ms with $P,Q>1$
have a larger set of high-gradient operators than 
in the projective (super) spaces. 
This can be seen by comparing the set of 
high-gradient operators~(\ref{eq: set of HGO})
against their counterparts when the target manifold is
$\mathrm{U}(P+Q)/\mathrm{U}(P)\times\mathrm{U}(Q)$
with $P,Q>1$, which can be found in
Eqs.\ (2.12) and (2.16b) from Ref.\ \onlinecite{Ryu07a}. 
High-gradient operators  for
$\mathrm{U}(P+Q)/\mathrm{U}(P)\times\mathrm{U}(Q)$
NL$\sigma$Ms with $P,Q>1$
can be expressed as a
product of traces of matrix fields, while 
in the complex projective space, there is no such trace. 
[Here, note that $\mathsf{Z}$ defined in Eq.\ (\ref{eq: Z and Y})
is a scalar while the corresponding object in 
$\mathrm{U}(P+Q)/\mathrm{U}(P)\times\mathrm{U}(Q)$
with $P,Q>1$ is a matrix, Eq.\ (2.9b) from Ref.~\onlinecite{Ryu07a}.]
Similarly, the set of high-gradient operators in the 
$\widehat{\mathrm{gl}}_{k>1}(M|M)$ WZW theory
(\ref{eq: def HGO if M,k to infty})
is larger than the set of high-gradient operators in 
the $\widehat{\mathrm{gl}}_{k=1}(M|M)$ WZW theory.

\subsection{
Sketch of the one-loop RG computation
           }

We now outline the calculations leading to 
the main results 
(\ref{eq: main result}) and (\ref{eq: main result2}). 

We choose the covariant background field method
to renormalize the NL$\sigma$Ms.
The merit of the background field method is
that there is no need to worry about the appearance of
redundant operators. This is very convenient when considering 
the mixing of a large set of operators under the RG,
that cannot be distinguished by the symmetries of the NL$\sigma$M.
The background field method consists in resolving the coordinates
$\varphi^{a}=\varphi^{a}_{\mathrm{cl}}+\zeta^{a}$
of a NL$\sigma$M into slow (mean-field) modes 
$\varphi^{a}_{\mathrm{cl}}$
that satisfy the classical equations of motion
and fast (fluctuating) modes $\zeta^{a}$
in terms of which the Taylor expansion
of the action transforms covariantly under 
reparametrization of the target manifold,
i.e., in terms of which only the metric,
the Riemann tensor, the covariant derivative, etc,
of the target manifold appear in the action.
For K\"ahler manifolds, this is achieved by choosing
$\zeta^{a}$ to be 
(either Riemannian or K\"ahlerian) normal coordinates.
The very same expansion of the action is also 
applied to the building blocks%
~(\ref{eq: def building blocks})
to the high-gradient operators, i.e., 
\begin{equation}
\mathcal{G}^{\ }_{\mu\nu}=
\left[\mathcal{G}^{\ }_{\mu\nu}\right]^{\ }_{\zeta^0}
+
\left[\mathcal{G}^{\ }_{\mu\nu}\right]^{\ }_{\zeta^1}
+
\cdots,
\end{equation}
where 
$\left[\mathcal{G}^{\ }_{\mu\nu}\right]^{\ }_{\zeta^p}$
represents a $p$-th term in this expansion.

To compute the anomalous scaling dimensions of high-gradient operators, 
they are first expanded in terms of the fast mode $\zeta^{a}$,
and are then pairwise Wick contracted. For example, 
the relevant formula for calculating 
$
\langle
\big[
\mathcal{G}_{\mu\nu}
\big]^{\ }_{\zeta^2}
\rangle
$
and
$
\langle
\big[
\mathcal{G}_{\mu\nu}
\big]^{\ }_{\zeta^1}
\big[
\mathcal{G}_{\rho\sigma}
\big]^{\ }_{\zeta^1}
\rangle
$,
whereby the angular bracket $\langle \cdots \rangle$
denotes pairwise Wick contraction of the fast modes $\zeta^{a}$,
can be found in Ref.~\onlinecite{Ryu07a},
e.g., Eq.~(C.40). When applied to the $\mathbb{C}P^{N+M-1|N}$ 
NL$\sigma$M, 
we obtain
\begin{equation}
\begin{split}
&
\langle 
\big[
\mathcal{G}^{\ }_{\mu\nu}
\big]^{\ }_{\zeta^2}
\rangle=
-
I M \delta^{\ }_{\mu,+\nu} 
\mathcal{G}^{\ }_{\mu\nu}, 
\\
&
\big\langle
\big[
\mathcal{G}^{\ }_{\mu\nu}
\big]^{\ }_{\zeta^1}
\big[
\mathcal{G}^{\ }_{\rho\sigma}
\big]^{\ }_{\zeta^1}
\big\rangle=
I
\Big(
\delta^{\ }_{\rho,-\nu}
-
\delta^{\ }_{\rho,-\mu}
-\
\delta^{\ }_{\sigma,-\nu}
+
\delta^{\ }_{\sigma,-\mu}
\Big)
\\
&
\hphantom{
\big\langle
\big[
\mathcal{G}^{\ }_{\mu\nu}
\big]^{\ }_{\zeta^1}
\big[
\mathcal{G}^{\ }_{\rho\sigma}
\big]^{\ }_{\zeta^1}
\big\rangle=
         }
\times 
\Big(
\mathcal{G}^{\ }_{\mu\sigma}
\mathcal{G}^{\ }_{\rho\nu}
+
\mathcal{G}^{\ }_{\mu\nu}
\mathcal{G}^{\ }_{\rho\sigma}
\Big),
\end{split} 
\end{equation}
where $I = \int d^d k/(2\pi)^d (1/k^2)$. 
After substituting $\mu=\pm,\nu=\pm$, this gives 
\begin{equation}
\begin{split}
&
\langle 
\big[
\mathcal{G}^{\ }_{++}
\big]^{\ }_{\zeta^2}
\rangle=
-
I M 
\mathcal{G}^{\ }_{++},
\\
&
\langle 
\big[
\mathcal{G}^{\ }_{--}
\big]^{\ }_{\zeta^2}
\rangle=
-
I M
\mathcal{G}^{\ }_{--},
\\
&
\big\langle
\big[
\mathcal{G}^{\ }_{++}
\big]^{\ }_{\zeta^1}
\big[
\mathcal{G}^{\ }_{++}
\big]^{\ }_{\zeta^1}
\big\rangle=
\big\langle
\big[
\mathcal{G}^{\ }_{-+}
\big]^{\ }_{\zeta^1}
\big[
\mathcal{G}^{\ }_{++}
\big]^{\ }_{\zeta^1}
\big\rangle
=
0,
\\
&
\big\langle
\big[
\mathcal{G}^{\ }_{+-}
\big]^{\ }_{\zeta^1}
\big[
\mathcal{G}^{\ }_{+-}
\big]^{\ }_{\zeta^1}
\big\rangle=
+4I
\mathcal{G}^{\ }_{+-}
\mathcal{G}^{\ }_{+-},
\\
&
\big\langle
\big[
\mathcal{G}^{\ }_{+-}
\big]^{\ }_{\zeta^1}
\big[
\mathcal{G}^{\ }_{-+}
\big]^{\ }_{\zeta^1}
\big\rangle=
-2I
\Big(
\mathcal{G}^{\ }_{++}
\mathcal{G}^{\ }_{--}
+
\mathcal{G}^{\ }_{+-}
\mathcal{G}^{\ }_{-+}
\Big),
\\
&
\big\langle
\big[
\mathcal{G}^{\ }_{-+}
\big]^{\ }_{\zeta^1}
\big[
\mathcal{G}^{\ }_{+-}
\big]^{\ }_{\zeta^1}
\big\rangle=
-2
I
\Big(
\mathcal{G}^{\ }_{--}
\mathcal{G}^{\ }_{++}
+
\mathcal{G}^{\ }_{-+}
\mathcal{G}^{\ }_{+-}
\Big), 
\\
&
\big\langle
\big[
\mathcal{G}^{\ }_{-+}
\big]^{\ }_{\zeta^1}
\big[
\mathcal{G}^{\ }_{-+}
\big]^{\ }_{\zeta^1}
\big\rangle=
+4I
\mathcal{G}^{\ }_{-+}
\mathcal{G}^{\ }_{-+}. 
\end{split}
\end{equation}
Furthermore, if $p$, $q$, and $r$ are 
non-negative integers, we find
\begin{widetext}
\begin{equation}
\begin{split}
\big\langle
\big[
\mathcal{G}_{+-}^{p}
\mathcal{G}_{-+}^{q}
\big]^{\ }_{\zeta^2}
\big\rangle 
=&\,
+
p\,
\big\langle
\big[
\mathcal{G}^{\ }_{+-} 
\big]^{\ }_{\zeta^2}
\big\rangle \,
\mathcal{G}_{+-}^{p-1}
\mathcal{G}_{-+}^{q}
\\
&\,
+
q\,
\mathcal{G}_{+-} ^{p}
\big\langle
\big[
\mathcal{G}^{\ }_{-+} 
\big]^{\ }_{\zeta^2}
\big\rangle 
\mathcal{G}_{-+}^{q-1}
\\
&\,
+
pq\,
\mathcal{G}_{+-}^{p-1}
\mathcal{G}_{-+}^{q-1}
\big\langle
\big[
\mathcal{G}^{\ }_{+-}
\big]^{\ }_{\zeta^1}
\big[
\mathcal{G}^{\ }_{-+}
\big]^{\ }_{\zeta^1}
\big\rangle 
\\
&\,
+
\frac{p(p-1)}{2}
\mathcal{G}_{+-}^{p-2}
\mathcal{G}_{-+}^{q}
\big\langle
\big[
\mathcal{G}^{\ }_{+-}
\big]^{\ }_{\zeta^1}
\big[
\mathcal{G}^{\ }_{+-}
\big]^{\ }_{\zeta^1}
\big\rangle 
\\
&\,
+
\frac{q(q-1)}{2}
\mathcal{G}_{+-}^{p}
\mathcal{G}_{-+}^{q-2}
\big\langle
\big[
\mathcal{G}^{\ }_{-+}
\big]^{\ }_{\zeta^1}
\big[
\mathcal{G}^{\ }_{-+}
\big]^{\ }_{\zeta^1}
\big\rangle 
\\
=&\,
+2I
\big[
-pq
+p(p-1)
+q(q-1)
\big]
\mathcal{G}_{+-}^{p}
\mathcal{G}_{-+}^{q}
-
2I pq\,
\mathcal{G}_{+-}^{p-1}
\mathcal{G}_{-+}^{q-1}
\mathcal{G}^{\ }_{++}
\mathcal{G}^{\ }_{--}
\end{split}
\end{equation}
and
\begin{equation}
\begin{split}
\langle \,
\big[
\mathcal{G}_{+-}^{p}
\mathcal{G}_{-+}^{q}
\big(
\mathcal{G}^{\ }_{++}
\mathcal{G}^{\ }_{--}
\big)^{r}
\big]^{\ }_{\zeta^2}
\,
\rangle
=&\,
+
r
\langle \,
\mathcal{G}_{+-}^{p}
\mathcal{G}_{-+}^{q}
\mathcal{G}_{++}^{r-1}
\big[
\mathcal{G}^{\ }_{++}
\big]^{\ }_{\zeta^2}
\mathcal{G}_{--}^{r}
\,\rangle
\\
&\,
+
r
\langle \,
\mathcal{G}_{+-}^{p}
\mathcal{G}_{-+}^{q}
\mathcal{G}_{++}^{r}
\mathcal{G}_{--}^{r-1}
\big[
\mathcal{G}^{\ }_{--}
\big]^{\ }_{\zeta^2}
\,\rangle
\\
&\,
+
\langle \,
\big[
\mathcal{G}_{+-}^{p}
\mathcal{G}_{-+}^{q}
\big]^{\ }_{\zeta^2}
\big(
\mathcal{G}^{\ }_{++}
\mathcal{G}^{\ }_{--}
\big)^{r}
\,
\rangle
\\
=&\,
-2I M r\,
\mathcal{G}_{+-}^{p}
\mathcal{G}_{-+}^{q}
\big(
\mathcal{G}^{\ }_{++}
\mathcal{G}^{\ }_{--}
\big)^{r}
\\
&\,
-2I\big[
pq
-p(p-1)
-q(q-1)
\big]
\mathcal{G}_{+-}^{p}
\mathcal{G}_{-+}^{q}
\big(
\mathcal{G}^{\ }_{++}
\mathcal{G}^{\ }_{--}
\big)^{r}
\\
&\,
-
2I pq\,
\mathcal{G}_{+-}^{p-1}
\mathcal{G}_{-+}^{q-1}
\big(
\mathcal{G}^{\ }_{++}
\mathcal{G}^{\ }_{--}
\big)^{r+1}. 
\end{split}
\label{eq: master formula for rg hgo cpn}
\end{equation}
\end{widetext}
Equation~(\ref{eq: master formula for rg hgo cpn})
justifies the claim that the family%
~(\ref{eq: set of HGO})
of high-gradient operators is closed 
under one-loop RG and yields Eqs.%
~(\ref{eq: main result}) and (\ref{eq: main result2}).

\section{ 
Relationship between the 
$\widehat{\mathrm{gl}}(2N|2N)^{\ }_{k=1}$ Thirring model
and Anderson localization with ``sublattice'' symmetry -- 
a review
        }
\label{sec: HWK model}

The $\widehat{\mathrm{gl}}(2N|2N)^{\ }_{k=1}$ Thirring model
represents the physics of observables in a class of problems of
Anderson localization in symmetry classes BDI
(see Ref.~\onlinecite{Guruswamy00}) and CII 
(see Ref.~\onlinecite{Ryu09})
within the classification scheme of
Refs.~\onlinecite{Verbaarschot94}--\onlinecite{Heinzner05}.
The fundamental physical observables are disorder averages of
(products) of Green's functions.
Here, we review some basic steps of this connection
for the example of symmetry class BDI, whose simplest
representative is a two-dimensional random tight-binding model
for fermions on a bipartite lattice. A popular
tight-binding model of that kind has recently become
that on a honeycomb lattice, 
due to its relevance for the physics of
graphene.
(Another, lattice realization of the same continuum physics
was obtained earlier in Refs.~\onlinecite{Hatsugai97} 
and \onlinecite{Mudry03}.)

\subsection{
Definitions
           }

To be specific, 
consider the low-energy properties of the tight-binding model on the
two-dimensional honeycomb lattice.  Only sites on the different sublattices
are connected by hopping, and the hopping matrix elements are independent
real random numbers with non-vanishing mean. Because of the constraint
that only sites  on different sublattices are connected, the model
inherits a special symmetry called sublattice (or chiral) symmetry. It
turns out to imply the presence of an operator that anticommutes 
with the Hamiltonian,
which thus relates the spectrum at positive and negative energies,
and makes the zero of energy, $E=0$ 
(often called the ``band center''), special.
Taking the low-energy limit near zero energy one obtains a random
Dirac equation. (See, e.g., Ref.~\onlinecite{Foster06} for details.)
We now start from the continuum limit of the so-obtained Hamiltonian,
which reads
\begin{subequations}
\label{eq: def random Dirac H}
\begin{equation}
\begin{split}
&
  \mathcal{H}
  =
  \mathcal{H}^{\ }_{0}
  +
  \mathcal{V}(\boldsymbol{r}),
\label{eq: HWK 1}
\end{split}
\end{equation}
where the kinetic energy is
(we set $\hbar$ and the Fermi velocity $v^{\ }_\mathrm{F}$ to be one)
\begin{equation}
\begin{split}
  \mathcal{H}^{\ }_{0}=
  -\sum_{\mu=1}^{2}{i}(\sigma^{\ }_\mu\otimes\tau^{\ }_1)\partial^{\ }_\mu,
\end{split}
\end{equation}
and the static disorder is
\begin{equation}
\begin{split} 
&
  \mathcal{V}(\boldsymbol{r})=
  \sum_{\mu=1}^{2}(\sigma^{\ }_\mu\otimes\tau^{\ }_2)A^{\ }_\mu(\boldsymbol{r}) 
  - 
  (\sigma^{\ }_0\otimes\tau^{\ }_2)V(\boldsymbol{r}) 
\\
&\hphantom{\mathcal{V}(\boldsymbol{r})=}
  + 
  (\sigma^{\ }_3\otimes\tau^{\ }_1)M(\boldsymbol{r}).
\end{split}
  \label{eq: HWK Hamiltonian }
\end{equation}
\end{subequations}
Here, $\sigma^{\ }_{1,2,3}$ and $\tau^{\ }_{1,2,3}$ 
are two independent sets of Pauli matrices together
with another two independent 
$2\times2$ identity matrix 
$\sigma^{\ }_{0}\equiv\mathbb{I}^{\ }_2$
and 
$\tau^{\ }_{0}\equiv\mathbb{I}^{\ }_2$.
The $2\times2$ matrix space associated  with
the $\tau$ Pauli matrices originates from the
bipartite symmetry of the underlying lattice model.
The real-valued functions (potentials)
$A^{\ }_{\mu}(\boldsymbol{r})$,
$V(\boldsymbol{r})$,
and
$M(\boldsymbol{r})$,
which do not vary appreciably on the scale of the lattice spacing,
represent four independent sources of (static) randomness.

The above potentials are random variables.
We will assume first that they are
white-noise distributed according to 
a Gaussian probability distribution with vanishing mean,
\begin{equation}
\begin{split}
&
  \overline{A^{\ }_{\mu}(\boldsymbol{r})A^{\ }_{\nu}(\boldsymbol{r}')}
  =
  g^{\ }_{\mathrm{A}} 
  \delta^{\ }_{\mu\nu}  
  \delta^{(2)}(\boldsymbol{r}-\boldsymbol{r}'),
  \quad \mu,\nu=1,2,
\\
&
  \overline{V(\boldsymbol{r})V(\boldsymbol{r}') }
  =
  \overline{M(\boldsymbol{r})M(\boldsymbol{r}') }
  =
  g^{\ }_{\mathrm{M}} \delta^{(2)}(\boldsymbol{r}-\boldsymbol{r}'),
\end{split}
\end{equation}
where 
$\delta^{(2)}(\boldsymbol{r}-\boldsymbol{r}')$ 
is the two-dimensional delta function,
$\overline{\cdots}$ represents disorder averaging, 
and we assume that the variances of 
$V(\boldsymbol{r})$ and $M(\boldsymbol{r})$ 
are identical.
Of course, the disorder strengths
$g^{\ }_{\mathrm{A}}$
and
$g^{\ }_{\mathrm{M}}$
are positive.

The tight-binding Hamiltonian 
is invariant under time-reversal and so is
its continuum limit
\begin{eqnarray}
  \mathcal{T}
  \left(\mathcal{H}\right)^*
  \mathcal{T}=
  \mathcal{H},
  \qquad
  \mathcal{T}:=
  \sigma^{\ }_1\otimes\tau^{\ }_3.
\label{TimeReversalSymmetry}
\end{eqnarray}
Since the tight-binding 
Hamiltonian preserves the bipartite nature of the underlying lattice
for any realization of the disorder, so does its continuum limit
through the chiral symmetry
\begin{eqnarray}
  \mathcal{C}\,
  \mathcal{H}\,
  \mathcal{C}=
  -\mathcal{H},
  \qquad
  \mathcal{C}:=
  \sigma^{\ }_0\otimes\tau^{\ }_3.
\end{eqnarray}
As already mentioned, because of its chiral and time reversal
symmetries  the Hamiltonian belongs to the so-called
BDI symmetry class 
within the classification scheme of
Refs.~\onlinecite{Verbaarschot94}--\onlinecite{Heinzner05}.

\subsection{ 
Path integral representation of the single-particle Green's function
           }

In problems of Anderson localization,
physical quantities are expressed by disorder averages of
(products of) the retarded and advanced Green's functions 
\begin{equation}
  \hat G^{{R}/{A}}(E):=
  (E\pm {i}\eta -\mathcal{H})^{-1}.
\end{equation}
In the present model, the retarded and advanced Green's functions 
are related, at the band center $E=0$, by the chiral symmetry through
\begin{eqnarray}
  \mathcal{C}\,\hat G^{{R}}(E=0)\,\mathcal{C}=
  -\hat G^{{A}}(E=0).
\end{eqnarray}
Hence, any arbitrary product of retarded or advanced Green's
function at the band center equates, up to a sign, 
a product of retarded Green's functions at the band center.
{}From now on we will omit the energy argument of the Green's
function having in mind that it is always fixed to the band center $E=0$.

Since the two kinds of Green's functions are related
by the chiral symmetry, it suffices to introduce functional
integrals for the retarded Green's function
defined by the supersymmetric partition function  
$Z\equiv Z^{\ }_{{F}}\times Z^{\ }_{{B}}$ with%
~\cite{Efetov97}
\begin{equation}
\begin{split}
&
  Z^{\ }_{{F}}:=
  \int\mathcal{D}[\bar{\chi},\chi]
  \exp   
  \left(
  {i}
  \int_r \,
  \bar{\chi}
  \left(
    {i}\eta
    -\mathcal{H}
  \right)
  \chi
  \right),
\\
&
  Z^{\ }_{{B}}:=
  \int\mathcal{D}[\bar{\xi},\xi]
  \exp   
  \left(
  {i}
  \int_r \,
  \bar{\xi}
  \left(
    {i}\eta
    -\mathcal{H}
  \right)
  \xi
  \right).
\end{split}
\end{equation}
Here,
$\int_r=\int d^2\boldsymbol{r}=\int d\bar zdz/(2{i})$,
$(\bar{\chi},\chi)$ is a pair of two independent 
four-component fermionic fields,
and
$(\bar{\xi},\xi)$ 
is a pair of four-component bosonic fields related by complex conjugation.

The matrix elements of the retarded Green's function
can be represented as
\begin{equation}
\begin{split}
{i}\hat{G}^{{R}}(\boldsymbol{r},\boldsymbol{r}')=&
  \langle
  \chi(\boldsymbol{r})\bar{\chi}(\boldsymbol{r}')
  \rangle
\\
  =&
  \langle
  \xi(\boldsymbol{r})\bar{\xi}(\boldsymbol{r}')
  \rangle
\end{split}
\label{eq: first rep G(r,r')}
\end{equation}
with $\langle\cdots\rangle$ denoting the expectation value taken with
the partition function $Z$. With the help of the property
$\mathcal{T}=\mathcal{T}^{\mathrm{T}}$ 
of time-reversal and the property in Eq.~(\ref{TimeReversalSymmetry}),
\begin{equation}
\begin{split}
&
\int_r
  \bar{\chi}
  ({i}\eta-\mathcal{H}) \chi=
  -
\int_r
  \chi^{\mathrm{T}}\mathcal{T} 
  \left({i}\eta-
    \mathcal{H}
  \right) 
  \mathcal{T} \bar{\chi}^{\mathrm{T}},
\\
&
\int_r
\bar{\xi}
  ({i}\eta-\mathcal{H}) \xi=
  +
\int_r
  \xi^{\mathrm{T}}\mathcal{T}
  \left({i}\eta-
    \mathcal{H}
  \right) 
  \mathcal{T} \bar{\xi}^{\mathrm{T}},
\label{eq: TR 1}
\end{split}
\end{equation}
Eq.~(\ref{eq: first rep G(r,r')})
is also given by
\begin{equation}
\begin{split}
{i} 
\hat{G}^{{R}}(\boldsymbol{r},\boldsymbol{r}')=&
  -
  \langle
  (\mathcal{T} \bar{\chi}^{\mathrm{T}})(\boldsymbol{r})
  (\chi^{\mathrm{T}}\mathcal{T})(\boldsymbol{r}')
  \rangle
\\
  =&
  +
  \langle
  (\mathcal{T} \bar{\xi}^{\mathrm{T}})(\boldsymbol{r})
  (\xi^{\mathrm{T}}\mathcal{T})(\boldsymbol{r}')
  \rangle.
\label{eq: TR 2}
\end{split}
\end{equation}

We now perform the change of integration variables 
$\bar{\chi},\chi \to \psi^{\dag},\psi$ and
$\bar{\xi},\xi \to \beta^{\dag},\beta$ where, 
when the matrix space on which the $\tau$ Pauli matrices
acts is made explicit,
\begin{equation}
\begin{split}
&
  \bar{\chi}
  \to
  \sqrt{
    \frac{1}{2\pi } }
  \left(
      \psi^{1\dag}\sigma^{\ }_x,  -{i}\psi_2^{\ }\sigma^{\ }_x
  \right),
\quad
  \chi
  \to
  \sqrt{
    \frac{1}{2\pi }} 
  \left(
    \begin{array}{c}
      +{i}\psi^{2\dag} \\
      \psi_1^{\ } 
    \end{array}
  \right),
\\
&
  \bar{\xi}
  \to
  \sqrt{
    \frac{1}{2\pi } }
  \left(
      \beta^{1\dag}\sigma^{\ }_x,  -{i}\beta_2^{\ }\sigma^{\ }_x
  \right),
\quad
  \xi
  \to
  \sqrt{
    \frac{1}{2\pi }} 
  \left(
    \begin{array}{c}
      -{i}\beta^{2 \dag} \\
      \beta_1^{\ } 
    \end{array}
  \right).
\end{split}
\end{equation}
We also define
\begin{equation}
\begin{split}
&
\bar{\mathsf{A}}:= A^{\ }_x+{i}A^{\ }_y,
\qquad
     \mathsf{A} := A^{\ }_x-{i}A^{\ }_y,
\\
&
\bar{m}:= V- {i}M,
\qquad
     m := V+ {i}M,
\end{split}
\end{equation}
With these changes of variables, 
the partition function at $E=0$ can be written as
\begin{subequations}
\label{eq: final partition fct HWK}
\begin{equation}
\begin{split}
&
Z^{\ }_F= 
  \int\mathcal{D}[\psi^{\dag},\psi]\
  e^{-S^{F}-S^{F}_{{i}\eta} 
    },
\\
&
Z^{\ }_B =
  \int\mathcal{D}[\beta^{\dag},\beta]\
  e^{
    -S^{B}-S^{B}_{{i}\eta}
    },
\end{split}
\end{equation}
with the effective action for the fermionic part given by
  \begin{eqnarray}
    S^{{F}}
    &=&
    \int_r\frac{1}{2\pi}
    \sum_{a=1}^{2}
    \Big[
    \psi^{a\dag}
    (2\bar{\partial}+\bar{\mathsf{A}})
    \psi^{ \   }_{a}
    +
    \bar\psi^{a \dag}
    (2     \partial +     \mathsf{A} )
    \bar\psi^{   \  }_{a}
\nonumber\\
&&
\qquad
\qquad
    +
    \bar{m}\psi^{a\dag}\bar\psi^{\ }_{a}
    +    
         m \bar\psi^{a\dag}\psi^{\ }_{a}
    \Big],
 \\
    S_{{i}\eta}^{{F}}
    &=&
\int_r
\frac{{i}\eta}{2\pi}
    \left(
      \psi^{1\dag}\bar\psi^{2\dag}
      +
      \bar\psi^{1\dag}\psi^{2\dag}
      -
      \psi^{\ }_{2}\bar\psi^{\ }_{1}
      -
      \bar\psi^{\ }_{2}\psi^{\ }_{1}
    \right),
\nonumber
  \end{eqnarray}
and the bosonic part of the effective action given by
  \begin{eqnarray}
    S^{{B}}
    &=&
    \left(
        \psi,\bar\psi \to \beta,\bar\beta,
        \quad 
        \psi^{\dag},\bar\psi^{\dag} \to \beta^{\dag},\bar\beta^{\dag}
        \hbox{ in $S^{{F}}$}
    \right),
\nonumber\\
&&\\
    S_{{i}\eta}^{{B}}
    &=&
   \int_r
\frac{  {i}\eta }{2\pi}
\left(
      -
      \beta^{1\dag}\bar\beta^{2\dag}
      -
      \bar\beta^{1\dag}\beta^{2\dag}
      -
      \beta^{\ }_{2}\bar\beta^{\ }_{1}
      -
      \bar\beta^{\ }_{2}\beta^{\ }_{1}
\right).
\nonumber
\end{eqnarray}
\end{subequations}
Observe the non-Hermitian appearance and
asymmetry between fermions and bosons in $S^{F/B}_{{i}\eta}$, 
which are necessary to maintain supersymmetry.%
~\cite{Guruswamy00,Mudry03}

The time-reversal invariance 
(\ref{eq: TR 1})
and
(\ref{eq: TR 2})
in terms of the new basis
implies invariance under%
~\cite{footnote: TRS and NLSM manifold}
\begin{equation}
\begin{split}
&
\psi_{2}^{\ } \to \psi_{1}^{\ },
\qquad
\psi_{1}^{\ } \to -\psi_{2}^{\ },
\\
&
\beta_{2}^{\ } \to -{i}\beta_{1}^{\ },
\qquad
\beta_{1}^{\ } \to +{i}\beta_{2}^{\ }.
\end{split}
\end{equation}

The finite level-broadening term $S^{F/B}_{{i}\eta}$ is necessary
for the computation of certain physical observables,
including for example the Kubo conductivity,
the Einstein conductivity, 
and the local density of states.%
\cite{footnote: Kubo vs Landauer}
However, when we compute the conductance from the 
Landauer formula by attaching ideal leads to 
the disordered region described by
the Hamiltonian (\ref{eq: HWK 1}),
we can set $\eta=0$ in the disordered region
(while still keeping $\eta\neq 0$ in the leads).%
~\cite{Ryu07b}

The last step consists 
in averaging the partition function 
$Z=Z^{\ }_{F}\times Z^{\ }_{B}$
over the probability distribution for the 
white-noise and Gaussian distributed random potentials.
In this way, one finds a generating function for 
the averages of Green's functions
which is nothing but the
$\widehat{\mathrm{gl}}(2|2)^{\ }_{k=1}$ Thirring model. 
Specially, integration over the vector potential yields 
the term proportional to
\begin{subequations}
\label{eq: integrating over disorder}
\begin{equation}
\begin{split}
\psi^{A\dag}
\psi^{\ }_{A}
\times
\bar\psi^{B\dag}
\bar\psi^{\ }_{B}
=&
(-)^{A}
\psi^{\ }_{A}
\psi^{A\dag}
\times
(-)^{B}
\bar\psi^{\ }_{B}
\bar\psi^{B\dag},
\end{split}
\end{equation}
while integration over the complex-valued mass yields 
the term proportional to
\begin{equation}
\begin{split}
    \psi^{B\dag}
\bar\psi^{\ }_{B}
\times
\bar\psi^{A\dag}
    \psi^{\ }_{A}
=&
(-)^{A}
    \psi^{\ }_{A}
    \psi^{B\dag}
\
\bar\psi^{\ }_{B}
\bar\psi^{A\dag},
\end{split}
\end{equation}
\end{subequations}
where we have combined bosonic 
$\beta^{\ }_a,\bar{\beta}^{\ }_a$ 
and fermionic 
$\psi^{\ }_a,\bar{\psi}^{\ }_a$ 
($a=1,2$) 
spinors into the supersymmetric vector
$\psi^{\ }_A, \bar{\psi}^{\ }_A$ ($A=1,\cdots,4$)
as in Sec.\ \ref{sec: gl(M|N) level k=1}.
The $N$-th moment of the retarded Green's function
evaluated at the band center is obtained by allowing the index 
$a$ to run from 1 to $2N$ in Eq.~(\ref{eq: final partition fct HWK})
or, equivalently, by allowing the indices $A$ and $B$ to run from 
1 to $2N+2N$ in Eq.~(\ref{eq: integrating over disorder}), 
thereby obtaining the
$\widehat{\mathrm{gl}}(2N|2N)^{\ }_{k=1}$ Thirring model.

We have assumed so far that the random imaginary vector potential
$\mathsf{A}, \bar{\mathsf{A}}$ and
the complex random mass $\bar{m},m$
possess a Gaussian probability distribution.
If we assume instead that their distributions have non-vanishing
higher cumulants (but still assuming that they have no spatial correlations),
the quenched disorder averaging necessarily yields
high-gradient operators of the form%
~(\ref{eq: hgo glMNk=1}).


\begin{thebibliography}{99}

\bibitem{Efetov97} 
For an extensive overview, see, e.g., K.\ Efetov, 
\textit{Supersymmetry in disorder and chaos}, 
(Cambridge University Press, Cambridge, 1997).

\bibitem{Mirlin00}
A. D. Mirlin, Phys.\ Rep.\ \textbf{326}, 259 (2000).

\bibitem{Foster09}
M. S. Foster, S. Ryu, and A. W. W. Ludwig, 
Phys.\ Rev.\ B \textbf{80}, 0705101 (2009);
see also \textit{Viewpoint} by
T. Vojta, Physics 2, 66 (2009).

\bibitem{Friedan85}
D. Friedan, 
Ann.~Phys.~(N.Y.) \textbf{163}, 318 (1985).

\bibitem{Lee85}
For a review, see P. A. Lee and T. V. Ramakrishnan,
Rev.~Mod.~Phys.~\textbf{57}, 287 (1985).

\bibitem{Kravtsov88} 
  V.\ E.\ Kravtsov, I.\ V.\ Lerner, and V.\ I.\ Yudson, 
  Sov.\ Phys.\ JETP \textbf{68}, 1441 (1988).

\bibitem{Kravtsov89} 
  V.\ E.\ Kravtsov, I.\ V.\ Lerner, and V.\ I.\ Yudson, 
  Phys.\ Lett.\ \textbf{134A}, 245 (1989).

\bibitem{Wegner90} 
  F.\ Wegner, Z.\ Phys.\ B \textbf{78}, 33 (1990).

\bibitem{Lerner90} 
  I.\ V.\ Lerner, and F.\ Wegner, Z.\ Phys.\ B \textbf{81}, 95 (1990).

\bibitem{Wegner91} 
  F.\ Wegner, Nucl.\ Phys.\ B \textbf{354}, 441 (1991).

\bibitem{Mall93} 
  H.\ Mall and F.\ Wegner, 
  Nucl.\ Phys.\ B \textbf{393}, 495 (1993).

\bibitem{Vasilev93}
A.\ N.\ Vasilev and A.\ S.\ Stepanenko,
Theor.\ Math.\ Phys.\ \textbf{94}, 471 (1993).

\bibitem{Lang94}
K.\ Lang and W.\ R\"uhl,
Z.\ Phys.\ C \textbf{61}, 495 (1994).

\bibitem{Derkachov97}
S.\ E.\ Derkachov, S.\ K.\ Kehrein, and A.\ N.\ Manashov,
Nucl.\ Phys.\ B \textbf{493}, 660 (1997).

\bibitem{Castilla97} 
  G.\ E.\ Castilla and S.\ Chakravarty, 
  Nucl.\ Phys.\ B \textbf{485}, 613 (1997).

\bibitem{Ryu07a}
S. Ryu, A. Furusaki, A. W. W. Ludwig, and C. Mudry,
Nucl.~Phys.~B \textbf{780}, 105 (2007).

\bibitem{footnote: our conventions for dimensions} 
We are using the following terminology.
A local operator $Q$ carries the scaling dimension $x^{\ }_{Q}$
if its two-point function decays algebraically fast 
for large separations with
the exponent $2x^{\ }_{Q}$,
$$
\langle Q(\boldsymbol{r})Q^{\dag}(0)\rangle\sim
|\mathfrak{a}/\boldsymbol{r}|^{2x^{\ }_{Q}}.
$$
If the critical theory is a free theory,
the scaling dimension $x^{\ }_{Q}$ coincides with
the engineering dimension $x^{(0)}_{Q}$ of $Q$.
The anomalous dimension $\gamma^{\ }_{Q}$ of the operator $Q$ is defined 
through 
$$
x^{\ }_{Q}:=
x^{(0)}_{Q}+\gamma^{\ }_{Q}.
$$

\bibitem{footnote: on anomalous dimensions}
We choose to refer to the corresponding RG function depending 
on the coupling constant(s) also as a \textit{scaling dimension} 
even if the coupling constants are not set
equal to a value corresponding to an RG fixed point.

\bibitem{Polyakov01}
  A.~M.~Polyakov,
  Int.\ J.\ Mod.\ Phys.\ A \textbf{18}, 1827 (2003).

\bibitem{Tseytlin03}
  A.~A.~Tseytlin,
  Nucl.\ Phys.\ B \textbf{664}, 247 (2003).

\bibitem{Polyakov05}
  A.~M.~Polyakov,
  arXiv:hep-th/0512310.

\bibitem{Ludwig90} 
  A.\ W.\ W.\ Ludwig, 
  Nucl.\ Phys.\ B \textbf{330}, 639 (1990).

\bibitem{Brezin97} 
  E.\ Brezin and S.\ Hikami, 
  Phys.\ Rev.\ B \textbf{55}, R10169 (1997).

\bibitem{Wess71}
J. Wess and B. Zumino,
Phys.~Lett.~\textbf{37B}, 95 (1971).

\bibitem{Novikov82}
S. P. Novikov, 
Usp.~Mat.~Nauk.~\textbf{37}, 3 (1982).

\bibitem{Witten84}
E.\ Witten,
Commun.\ Math.\ Phys.\ \textbf{92}, 455 (1984).

\bibitem{Polyakov83}
A. M. Polyakov and P. B. Wiegmann,
Phys.~Lett.~\textbf{131B}, 121 (1983).

\bibitem{Bocquet00}
for a review of non-Abelian Bosonization, 
Ref. \onlinecite{Witten84}, in the supersymmetric
case, see, e.g.,
M. Bocquet, D. Serban, and M. R. Zirnbauer, 
Nucl.\ Phys.\ B \textbf{578}, 628 (2000).


\bibitem{Altshuler86a} 
  B.\ L.\ Altshuler, V.\ E.\ Kravtsov, and I.\ V.\ Lerner, 
  JETP Lett.~\textbf{43}, 440 (1986).

\bibitem{Altshuler86b} 
  B.\ L.\ Altshuler, V.\ E.\ Kravtsov, and I.\ V.\ Lerner, 
  Sov.\ Phys.\ JETP \textbf{64}, 1352 (1986).

\bibitem{Altshuler89}  
  B.\ L.\ Altshuler, V.\ E.\ Kravtsov, and I.\ V.\ Lerner, 
  Phys.\ Lett.\ \textbf{134A}, 488 (1989).

\bibitem{Altshuler91} 
B.\ L.\ Altshuler, V.\ E.\ Kravtsov, and I.\ V.\ Lerner, 
in \textit{Mesoscopic Phenomena in Solids}, 
edited by B.\ L.\ Altshuler, P.\ A.\ Lee, and R.\ A.\ Webb, 
(North-Holland, Amsterdam, 1991), p.~449.

\bibitem{Guruswamy00} 
S.\ Guruswamy, A.\ LeClair, and A.\ W.\ W.\ Ludwig, 
Nucl.\ Phys.\ B \textbf{583}, 475 (2000).

\bibitem{Foster06}
M. S. Foster and A. W. W. Ludwig, 
Phys.\ Rev.\ B \textbf{73}, 155104 (2006).


\bibitem{Verbaarschot94}
J. J. M. Verbaarschot, 
Phys.\ Rev.\ Lett.\ \textbf{72}, 2531 (1994).

\bibitem{Zirnbauer96} 
M.\ R.\ Zirnbauer, 
J.\ Math.\ Phys.\ \textbf{37}, 4986 (1996).

\bibitem{Altland97} 
A.\ Altland and M.\ R.\ Zirnbauer, 
Phys.\ Rev.\ B \textbf{55}, 1142 (1997).

\bibitem{Heinzner05}
P. Heinzner, A. Huckleberry, and M. R. Zirnbauer,
Commun.\ Math.\ Phys.\ \textbf{257}, 725 (2005).

\bibitem{Hatsugai97} 
Y.\ Hatsugai, X.-G.\ Wen, and M.\ Kohmoto, 
Phys.\ Rev.\ B \textbf{56}, 1061 (1997).

\bibitem{Schnyder08}
A. P. Schnyder, 
S. Ryu, 
A. Furusaki, 
and
A. W. W. Ludwig,
Phys.~Rev.~B \textbf{78}, 195125 (2008);
A. P. Schnyder, S. Ryu, and A. W. W. Ludwig, arXiv:0901.1343
[cond-mat.mes-hall].

\bibitem{Ryu09}
S. Ryu, C.  Mudry, A. Furusaki, and A. W. W. Ludwig,
unpublished.

\bibitem{Hosur09} 
Pavan Hosur, Shinsei Ryu, and Ashvin Vishwanath,
\texttt{arXiv:0908.2691}. 


\bibitem{Affleck87}
I.\ Affleck and F.\ D.\ M.\ Haldane,
Phys.\ Rev.\ B \textbf{36}, 5291 (1987).

\bibitem{Shankar90}
R.\ Shankar and N.\ Read,
Nucl.\ Phys.\ B \textbf{336}, 457 (1990).


\bibitem{Zamolodchikov92}
A. B. Zamolodchikov and Al.\ B. Zamolodchikov, 
Nucl.\ Phys.\ B \textbf{379}, 602 (1992).


\bibitem{footnote: Einstein convention}
Summation over repeated contravariant and covariant indices is implied.



\bibitem{footnote: invariant tensor}
For a Lie group $G$ and its Lie algebra $\mathcal{L}(G)$,
the latter spanned by the generators $\{X^{\ }_i\}$
with the commutators
$
\left[X^{\ }_i,X^{\ }_j\right] = f^{\ }_{ijk}X^{\ }_k
$,
a tensor that satisfies 
$$
T^{\ }_{i_1 \cdots i_P}
=
U^{\ }_{i_1 j_1}
\cdots
U^{\ }_{i_P j_P}
T^{\ }_{j_1 \cdots j_P}
$$
for any $U$ in the adjoint representation, i.e.,
$U^{\ }_{jk}= \exp\left(- t^{\ }_i f^{\ }_{ijk} \right)
$
($t^{\ }_{i}\in \mathbb{R}$),
is called an invariant tensor for the adjoint representation.
The scalar constructed from the invariant tensor
$T^{\ }_{i_1 \cdots i_P}$
and the Lie algebra generators $\{X^{\ }_i\}$
according to 
$$
C^{(P)} = 
T^{\ }_{i_1 \cdots i_P}
X^{\ }_{i_1}\cdots X^{\ }_{i_P}
$$
is a $P$-th order Casimir, i.e., 
$[C^{(P)}, X^{\ }_i]=0$ for all $X^{\ }_i$.
For example, 
if $V(X^{\ }_i)$ is a representation of $X^{\ }_i$,
then
$$
\mathrm{tr}\,\left[
V(X^{\ }_{i_1})\cdots V(X^{\ }_{i_P})
\right]
$$
is an invariant tensor for the adjoint representation.
For more details, see
A.\ O.\ Barut and R.\ Raczka,
\textit{Theory of Group Representations and
Applications},
(Polish Scientific Publishers, Warszawa, 1977).


\bibitem{Dittner72}
P.\ Dittner,
Commun.\ Math.\ Phys.\ \textbf{27}, 44 (1972).

\bibitem{DiFrancesco97}
For a review, see e.g. A. W. W. Ludwig {\it in}
``Low-Dimensional Quantum Field Theories for Condensed
Matter Physicists'', eds. S. Lundqvist, G. Morandi, Yu Lu (World Scientific,
River Edge, 1995);
or:
Sec.\ 6.5 in 
P.\ Di Francesco, P.\ Mathieu, and D.\ S\'en\'echal
\textit{Conformal Field Theory},
(Springer-Verlag, New York, 1997).


\bibitem{footnote: choice HGO for s<k}
The integer $[s/2]$ is the largest integer
which is smaller or equal
to $s/2$. 


\bibitem{Bershadsky99} 
M.\ Bershadsky, S.\ Zhukov, and A.\ Vaintrob,
Nucl.\ Phys.\ B \textbf{559}, 205 (1999).

\bibitem{Berkovits99}
N.\ Berkovits, C.\ Vafa, and E.\ Witten,
J.\ High Energy Phys.\ \textbf{03}, 018 (1999).

\bibitem{Tsvelik07}
The $\widehat{\mathrm{psl}}(M|M)^{\ }_{k}$ WZW model
has also been studied as a candidate 
to describe the integer quantum Hall
plateau transition;
M. R. Zirnbauer, arXiv:hep-th/9905054;
M. J. Bhaseen, I. I. Kogan, O. A. Solovyev, N. Taniguchi,
and A. M. Tsvelik,
Nucl.\ Phys.\ B \textbf{580}, 688 (2000);
A. M. Tsvelik, 
Phys.\ Rev.\ B \textbf{75}, 184201 (2007).
-- See however Ref.~\onlinecite{Obuse08}.

\bibitem{Obuse08} 
H. Obuse, A. R. Subramaniam, A. Furusaki, I. A. Gruzberg, 
and A. W. W. Ludwig,
Phys.\ Rev.\ Lett.\ \textbf{101}, 116802 (2008).

\bibitem{Mudry96}
C. Mudry, C.\ Chamon, and X.-G. Wen, 
Nucl.\ Phys.\ B \textbf{466}, 383 (1996).

\bibitem{Mudry03} 
C.\ Mudry, S.\ Ryu, and A.\ Furusaki, 
Phys.\ Rev.\ B \textbf{67}, 064202 (2003).

\bibitem{Korshunov93}
S.\ E.\ Korshunov, 
Phys.\ Rev.\ B \textbf{48}, 1124 (1993);
T.\ Nattermann, S.\ Scheidl, S.\ E.\ Korshunov, and M.\ S.\ Li, 
J.\ Phys.\ I (France) \textbf{5}, 565 (1995);
S.\ E. Korshunov and T. Nattermann, 
Phys.\ Rev.\ B \textbf{53}, 2746 (1996).

\bibitem{Scheidl97}
S. Scheidl, 
Phys.\ Rev.\ B \textbf{55}, 457 (1997).

\bibitem{Mudry99}
C.\ Mudry and X.-G.\ Wen, 
Nucl.~Phys.~B \textbf{549}, 613 (1999).

\bibitem{Carpentier00} 
D.\ Carpentier and P.\ Le Doussal, 
Nucl.\ Phys.\ B \textbf{588}, 565 (2000).

\bibitem{Fukui02}
T.\ Fukui, 
Phys.\ Rev.\ B \textbf{68}, 153307 (2003).

\bibitem{LeDoussal06}
P.\ Le Doussal and G.\ Schehr,
Phys. Rev. B \textbf{75}, 184401 (2007).

\bibitem{Ostrovsky06}
P.\ M.\ Ostrovsky, 
I.\ V.\ Gornyi, 
and A.\ D.\ Mirlin,
Phys.\ Rev.\ B \textbf{74}, 235443 (2006).




\bibitem{Gepner86}
In this paper, we relied heavily on the free-fermion
realization of the current algebra to select the 
``quantum'' high-gradient operators
from the ``classical'' ones, i.e., to implement 
normal ordering in a way consistent with the current
algebra. However, it is of course not necessary to rely on the
free-fermion realization of the current algebra
to achieve this task.
[See, e.g., D. Gepner and E. Witten, 
Nucl.\ Phys.\ B \textbf{278}, 493 (1986).]



\bibitem{CommentSchomerusSaleur2009} Boundary properties of the
$\mathbb{C}P^{1|2}$ NL$\sigma$M have recently been studied
in Ref.~\onlinecite{SchomerusSaleurEtAl2009}.

\bibitem{SchomerusSaleurEtAl2009} 
C. Candu, V. Mitev, T. Quells, H. Saleur, and V. Schomerus,
\texttt{arXiv:0908.0878.v2}. 


\bibitem{SuperCPN}  
NL$\sigma$Ms on these supersymmetric spaces have also been discussed
in Refs.~\onlinecite{SchomerusSaleurEtAl2009,Read01}.


\bibitem{Read01}
N.\ Read, and H.\ Saleur,
Nucl.\ Phys.\ B \textbf{613}, 409  (2001). 


\bibitem{Ryu07b}
S. Ryu, C.  Mudry, A. Furusaki, and A. W. W. Ludwig,
Phys.\ Rev.\ B \textbf{75}, 205344 (2007).



\bibitem{Hof86}
D.\ H\"of and F.\ Wegner,
Nucl.\ Phys.\ B \textbf{275}, 561 (1986).

\bibitem{Gade91-93} 
R.\ Gade and F.\ Wegner,
Nucl.\ Phys.\ B \textbf{360}, 213 (1991);
R.\ Gade, 
\textit{ibid.} \textbf{398}, 499 (1993). 






\bibitem{footnote: TRS and NLSM manifold}
As common when deriving a NL$\sigma$M
for a problem of Anderson localization with TRS,
the TRS can be made explicit, if desired, 
by doubling the number of components for the fields.

\bibitem{footnote: Kubo vs Landauer}
We refer the reader to Refs.~\onlinecite{Ludwig94} 
and~\onlinecite{Ryu07b} for the rationale to
distinguish the Kubo from the Einstein conductivity 
in the context of Dirac fermions.

\bibitem{Ludwig94} 
A.\ W.\ W.\ Ludwig, M.\ P.\ A.\ Fisher, R.\ Shankar, and G.\ Grinstein, 
Phys.\ Rev.\ B \textbf{50}, 7526 (1994).
 
\end{thebibliography}
\end{document}